\documentclass[prd,twocolumn,superscriptaddress,floatfix,amsmath,amssymb,amsfonts,nofootinbib]{revtex4-2}
\usepackage{float} % Paquete que permite la orden [H] = Pon esta figura aqu\'{i} bajo mi responsabilidad por muy feo que quede

\usepackage{comment}
\usepackage[normalem]{ulem}
\usepackage[english]{babel}
\usepackage{graphicx}% Include figure files
\usepackage{dcolumn}% Align table columns on decimal point
\usepackage{bm}% bold math
\usepackage{blindtext}
\usepackage{verbatim}
\usepackage{mathrsfs}
\usepackage{musicography}
\usepackage{amsmath}
\usepackage{dsfont}
\usepackage{cancel}
\usepackage{physics}
\usepackage{epstopdf}
\usepackage{mathtools}
\usepackage{color}
\usepackage[usenames,dvipsnames]{pstricks}
\usepackage{epsfig}
\usepackage{pst-grad} % For gradients
\usepackage{pst-plot} % For axes
\usepackage{hyperref}
\usepackage{verbatim}
\usepackage{afterpage}
\usepackage{tikz-feynman}
\hypersetup{
    colorlinks=true,
    linkcolor=blue,
    filecolor=magenta,      
    urlcolor=cyan,
}

\newcommand{\mf}{\mathsf}%\mathsf is too long

\newcommand{\ii}{\mathrm{i}}

\DeclareMathOperator*{\sumint}{%
\mathchoice%
  {\ooalign{$\displaystyle\sum$\cr\hidewidth$\displaystyle\int$\hidewidth\cr}}
  {\ooalign{\raisebox{.14\height}{\scalebox{.7}{$\textstyle\sum$}}\cr\hidewidth$\textstyle\int$\hidewidth\cr}}
  {\ooalign{\raisebox{.2\height}{\scalebox{.6}{$\scriptstyle\sum$}}\cr$\scriptstyle\int$\cr}}
  {\ooalign{\raisebox{.2\height}{\scalebox{.6}{$\scriptstyle\sum$}}\cr$\scriptstyle\int$\cr}}
}

\begin{document}

\title{Measurements in QFT:\\ Weakly coupled local particle detectors and entanglement harvesting}

\author{Daniel Grimmer}
\email{daniel.grimmer@philosophy.ox.ac.uk}
\affiliation{University of Oxford, Pembroke College, St. Aldates, Oxford, OX1 1DW, United Kingdom}

\author{Bruno de S. L. Torres}
\email{bdesouzaleaotorres@perimeterinstitute.ca}
\affiliation{Perimeter Institute for Theoretical Physics, Waterloo, Ontario, N2L 2Y5, Canada}
\affiliation{Department of Physics and Astronomy, University of Waterloo, Waterloo, Ontario, N2L 3G1, Canada}
\affiliation{Institute for Quantum Computing, University of Waterloo, Waterloo, Ontario, N2L 3G1, Canada}

\author{Eduardo Mart\'{i}n-Mart\'{i}nez}
\email{emartinmartinez@uwaterloo.ca}
\affiliation{Department of Applied Mathematics, University of Waterloo, Waterloo, Ontario, N2L 3G1, Canada}
\affiliation{Perimeter Institute for Theoretical Physics, Waterloo, Ontario, N2L 2Y5, Canada}
\affiliation{Institute for Quantum Computing, University of Waterloo, Waterloo, Ontario, N2L 3G1, Canada}

\begin{abstract}
We present a comparison of the AQFT-based Fewster-Verch framework  with the Unruh-DeWitt particle detector models commonly employed in relativistic quantum information and QFT in curved space. We use this comparison to respond to a recent paper \cite{ruep2021} in which it was argued that the Reeh-Schlieder theorem prevents weakly coupled local particle detectors from harvesting vacuum entanglement from a quantum field. Their claim can be traced back to the mixedness that a local particle detector cannot escape from because of the entanglement that it must have with the quantum fields outside of the localization area. We argue that for any realistic scale of localization for physical particle detectors the effect of that mixedness is negligible, and that weakly coupled localized particle detectors are not impeded from harvesting vacuum entanglement.

%We study the ability for local probes to harvest entanglement from the vacuum of a quantum field, when the initial states for the probes are mixed. We show that for sufficiently small coupling constant between probes and field, there is a threshold of initial mixedness beyond which no entanglement can be harvested---or equivalently, for a given initial mixedness, there is a minimum value for the coupling constant such that there is entanglement harvesting. We are able to verify this threshold both in cavities and in free space, modeling the detectors as localized harmonic oscillators.

\end{abstract}

\maketitle

\section{Introduction}
The power of quantum information as tool to formulate and tackle problems in fundamental physics is a subject that has drawn increasing levels of attention from a variety of areas in theoretical physics in recent years. This has led to the birth of various research programs that lie on the interface between quantum information and other well established areas of physics, including condensed matter, high energy physics, quantum field theory in curved spacetimes, and quantum gravity. One of the fields that recently emerged from such interplay is known as relativistic quantum information (RQI). Relativistic quantum information has the double-edged interest of using quantum information to understand better the structure of spacetime and the gravitational interaction, as well as studying information-theoretic tasks and concepts in setups where relativistic effects are relevant. RQI has produced results ranging from quantum optics and the light-matter interaction in relativistic settings~\cite{Richard},  to renewed perspectives on fundamental phenomena such as the Unruh and Hawking effects. 

One particular topic that permeates several developments in RQI involves the basic question of how the acquisition of information from a quantum system (more specifically, a relativistic quantum field) takes place. This process, in turn, is generally phrased in terms of how \emph{measurements} are to be modeled and conceptualized. Such an issue becomes especially relevant in relativistic setups, where additional subtleties associated with requirements of relativistic nature (such as locality, causality and general covariance in relativistic field theories) cannot be overlooked~\cite{Sorkin}. In RQI, the general approach to deal with this problem is formulated in terms of \emph{particle detector models}, which are localized, nonrelativistic quantum-mechanical probes which couple to quantum fields. 

The paradigmatic example of a particle detector model consists of a two-level (qubit) system that couples to a scalar field, in what has become known as the Unruh-DeWitt (UDW) detector. More generally, one can find several other examples of localized quantum mechanical systems that have been employed as probes to quantum fields (such as particles confined in a box or harmonic oscillators, e.g. ~\cite{Unruh1976, BLHu2007, Brown2013, Hotta2020, Zeromode} among others). Such examples are often referred to collectively as UDW-like models. Information from the field in this framework is then extracted indirectly by operations performed on the nonrelativistic probe system, where (for all practical purposes) the well-established measurement tools that we are familiar with from nonrelativistic quantum mechanics apply. This strategy is particularly well-suited at connecting fundamental questions in quantum field theory (in possibly curved spacetimes) with setups that can, in principle, be emulated in tabletop experiments. It has also been remarkably successful at providing operationally meaningful ways of probing more sophisticated quantum aspects of relativistic field theories, such as its entanglement structure~\cite{Valentini1991, Reznik2003, Pozas-Kerstjens:2015} and how it depends on properties of the background spacetime such as its geometry and topology~\cite{Menicucci, Terno2016, Cosmo, Henderson2018}. This is achieved through a process known as~\emph{entanglement harvesting}, where localized detectors can become entangled through their interaction with a quantum field (even when their coupling regions are spacelike separated) due to the fact that the initial state of the field itself is entangled.

A renewed interest in the question of how to model local measurements in quantum fields has recently emerged. This interest was motivated by the desire to formulate a measurement framework that is, by construction, compatible with our most rigorous understanding of quantum field theory in curved spacetimes. Such a measurement framework has been developed by Fewster and Verch~\cite{fewster1,fewster2}, and is phrased in the language of algebraic quantum field theory (AQFT). It establishes an important complementary perspective on the measurement problem, with emphasis on the fundamental requirements of relativistic locality and causality.

Recently, the Fewster-Verch (FV) framework was employed to provide a concrete toy theory---which we refer to as the \emph{Ruep model}~\cite{ruep2021}---of local probes consisting of localized modes of free scalar fields that couple to the target system of interest, with particular focus on the task of entanglement harvesting. The author of \cite{ruep2021} shows that if one uses such local modes as probes, obtained by restricting a Reeh-Schlieder state of a quantum field to a localized region, their initial state will be unavoidably mixed. This result is a consequence of the Reeh-Schlieder property and should be contrasted with what is usually assumed in entanglement harvesting scenarios in RQI, where detectors are (typically) initialized in pure states. This important difference---as well as its consequences to tasks such as entanglement harvesting---motivates a more thorough investigation on the contrast between the UDW and FV approaches to measurements in QFT, which is what we aim to address in this paper. 

Our analysis will show that, even though it is true that any local mode of a free scalar field in a Reeh-Schlieder state will be mixed, it is always possible to find localized modes that are arbitrarily close to perfect purity. Moreover, just characterizing a localization scale (given by the size of the region where the mode is supported) is not enough if one intends to assess the impact of the mixedness of local field modes when used as detectors. Shape matters as much as size. 

Our analysis will also show that local modes of a free field theory do not constitute reliable models for localized probes such as the ones inspired by setups found in quantum optics or atomic physics (which are the main motivation for UDW detector models). Therefore, despite establishing important milestones for the development of concrete toy models where the Fewster-Verch framework can be explored, probes modeled through localized modes of free scalar fields are likely to fall short when it comes to studying realistic setups. We also demonstrate, within the qubit and harmonic-oscillator UDW models, that the level of mixedness in a experimentally realistic regime can be controlled to an extent that effectively allows for entanglement harvesting with perturbatively small coupling.

The paper is organized as follows. Sec.~\ref{RQMProblem} contains an in-depth discussion on the measurement problem in quantum theory, both in nonrelativistic and relativistic contexts. We draw attention to a conceptually important distinction that can be made between the \mbox{Unruh-DeWitt} and Fewster-Verch approaches to modeling measurements in quantum fields. Namely, how one models the probe system in regards to what we will call the~\emph{relativistic cut}. Sec.~\ref{twodetectormodels} reviews the main technical ingredients in both the UDW-like models and the FV framework, as well as the main claims in~\cite{ruep2021} concerning the Ruep model of entanglement harvesting via local modes of a scalar field. In Sec.~\ref{RealisticMixedness} we show that, for any region size and ambient field temperature, there is always a spatial profile for the field mode such that its reduced state is arbitrarily pure. Furthermore, depending on what particular shape is chosen, the purity's dependence on the region size and ambient field temperature can qualitatively change. This shows that, without a more detailed motivation for a particular choice of local mode, the impact of more specific conclusions from such simple scalar field probes to actual entanglement harvesting scenarios is limited. Finally, Sec.~\ref{UDWperturbative} analyzes the result proven in~\cite{ruep2021} that the mixedness is ultimately responsible for a finite threshold for the value of the coupling strength below which no entanglement is harvested perturbatively. For the cases of the qubit and harmonic-oscillator UDW models, we show how one can derive, at lowest order in perturbation theory, a rather simple relation between the probes' initial mixedness and the entanglement harvesting degradation as compared to the case when the probes are initialized in pure states. This helps us give context to the claims made in~\cite{ruep2021} about the impossibility of entanglement harvesting at low coupling, by precisely quantifying what the threshold for the coupling strength is once a given initial level of mixedness is present.

\section{Background: The Quantum Field Measurement Problem}\label{RQMProblem}
Before we delve into a comparative analysis of the Ruep model and the more standard Unruh-DeWitt approach commonly employed in relativistic quantum information, we will  briefly review in this section the quantum measurement problem in both nonrelativistic and relativistic settings. As we will discuss, the primary difference between the Unruh-DeWitt and the Fewster-Verch framework (in which the Ruep model is framed) is in the way they approach the modeling of measurements of quantum fields. Particularly, the difference resides in where they place a certain quantum field theoretic version of the Heisenberg cut.

\subsection{The non-relativisitic quantum measurement problem}
There is a long-standing fundamental question at the heart of nonrelativistic quantum mechanics, commonly dubbed \textit{The Quantum Measurement Problem}: How are we to understand/model the measurement process which connects the quantum states in our laboratories to a string of classical data constituting the outcomes of each single-shot measurement? 

It is important to differentiate two aspects to the question, one pragmatic and one philosophical. The pragmatic question centers around how this measurement process (which goes from quantum probability distributions to definite  outcomes of single-shot experiments) should be \textit{modeled} mathematically. By contrast, the philosophical question centers around how this measurement process should be \textit{understood}/\textit{interpreted}. There is a wide variety of approaches to dealing with the philosophical aspect of the measurement problem, see \cite{sep-qt-issues} and references therein. However, a discussion of these is far outside of the scope of this paper. Our focus here is on the pragmatic modeling question.

Fortunately, in the nonrelativistic setting there is little disagreement about the pragmatic portion of the quantum measurement problem. We all agree about the relevant principles/notions: Born's rule~\cite{Ballentine}, L\"uders rule~\cite{Luders}, projector valued measurements (PVMs), positive operator-valued measurements (POVMs), nonselective measurements, postselective measurements, etc. Moreover, we all agree as to when and how they are to be applied. As a result, it is widely agreed how these rules are to be applied in a nonrelativistic context to arrive at the probabilities of postmeasurement outcomes. These methods of modeling nonrelativistic quantum measurements have been proven accurate enough by more than a century of experiments. Where there is disagreement about the quantum measurement problem, it is on the philosophical side, about what is \textit{really} going on here behind the scenes, so to speak.

An important, still open, question in the nonrelativistic measurement problem is the placement of the \textit{Heisenberg cut}. For our purposes\footnote{We make no claim that the way that we are using the term ``Heinsenberg cut'' here has any connection with Heinsenberg's understanding of quantum theory. Rather, we employ the term as it is used colloquially in modern discussions. No exegesis is attempted or implied.} we define this as the point where we can reasonably stop modeling the measurement process as being quantum. Below the cut the dynamics is modeled as a wavefunction evolving under the Schr\"{o}dinger equation; above the cut the dynamics is modeled classically. It should be stressed that as we are using this term, the Heisenberg cut is a theoretical construct. Indeed, we believe the world to be quantum through-and-through. Past the cut, we are no longer \textit{modeling} the measurement apparatus quantum mechanically; This is very different from the measurement apparatus no longer \textit{being} quantum past the Heisenberg cut. This cut can certainly be placed somewhere: an extreme example would be to say that we do not need quantum mechanics to model the notepad storing the information which the experimentalist gathered from their experiment, notwithstanding the fact that the notepad is quantum at the fundamental level.

To illustrate this example, suppose that an atom interacts with a microscopic detector which activates a transistor which starts a cascade of processes which eventually results in a display showing a number on a screen. The experimenter then sees this and writes it down on their notepad. For all practical purposes, we may take the Heisenberg cut at any point in the above sequence after about the transistor. This is because, after that point, enough decoherence has occurred that the possibility of spontaneous wide-scale recoherence (although not mathematically impossible) is practically inconceivable. That is, for modeling purposes it does not matter where we put the Heisenberg cut, so long as it is at a scale where quantum effects are (and will forever remain) irrelevant in practice. %\tcr{The nonrelativistic measurement theory discussed above is what one finds taking by introducing any late-enough Heisenberg cut.}

The above ``and will forever remain'' caveat is a critically important one. It reinforces the warning that the Heisenberg cut should not be thought of as being fundamental. Indeed, the validity of any Heisenberg cut approximation will always depend on the context surrounding the measurement procedure under consideration. In particular, one cannot simply decide mid-measurement to take a Heisenberg cut without knowing beforehand what the rest of the measurement procedure will be like. No matter how small quantum coherence effects appear to be in the middle of an experiment, there is always a possibility that the coherence effects are brought back to their full force\footnote{Such a carefully orchestrated large-scale recoherence is, in fact, exactly what quantum computers are designed to do.}, e.g., if the evolution is reversible. %Moreover, even if within one experiment the quantum coherence effects never again become relevant, they may become again relevant in other future measurements involving correlated systems. %The consideration of measurements made by observers who themselves live inside of a giant quantum computer capable of wide-scale (observers included) recoherence, leads to interesting Wigner's friend-like puzzles.

At this point one may wonder: if the application of a Heisenberg cut is a matter of nonfundamental pragmatic concern only, then do we really need it to make sense of the quantum measurement problem? One may ask: If we believe that the world is quantum through-and-through, then why would it be necessary to connect our quantum model of reality with an (incorrect) classical model of reality in order to understand it? Can we have a quantum-native understanding of quantum measurements? To answer these questions we need to distinguish between what is fundamental in a theory and what is foundational.

Given a theoretical framework, we understand as ``fundamental’’ what the theory says there is in the world, and how the theory says those things evolve and interact. For nonrelativistic quantum theory this is the wavefunction and the Schr\"odinger equation. According to this theory, the world exists and evolves in these terms only. The world takes no notice of which parts of its evolution might be characterized as a measurement, or where it might be valid to apply a Heisenberg cut. In this limited sense, the quantum world is understandable without a Heisenberg cut (and indeed without a measurement theory at all).%\footnote{\Dan{I am tempted to add a footnote here mentioning that Bohmians have their way of rewriting these equations to suggest a different ontology.}}

However, we argue that a theory is not completely understood until we have done some further foundational (possibly nonfundamental) work. In particular, our theory is not completely understood until we have identified within it who the observers are and what mathematical operations their measurements/experiments correspond to. At its heart, this is what the quantum measurement problem is about. Where are the observers in the theory and how they come to know the (apparently) single-valued outcomes of their experiments? Even if the observers are to be identified from the fundamental theory only in some emergent or approximate regime, the complexity of the emergent phenomena still enforces the need to address these foundational questions from a nonfundamental perspective. Therefore, we need to build a bridge between the fundamental aspects of the theory to the emergent or approximate theory in which our observers ``live''. For the case of quantum theory, this bridging involves developing an understanding of Heisenberg cuts so as to connect the quantum world with our preexisting classical-native understanding of observers and measurement outcomes.

In the following subsection, we will apply some of these lessons to the quantum field measurement problem. In particular, we will identify the need for a relativistic version of the Heisenberg cut in the modeling of measurements of quantum fields.

%\tcr{For the case of quantum theory, answering these questions will involve either a) developing a quantum-native understanding of observers and measurement outcomes, or b) developing and understanding of Heisenberg cuts to connect the quantum world with our existing classical-native understanding of observers and measurement outcomes. Since we, as humans, have evolved our perception and mental processing in a very-nearly classical world, the second option appears far superior.}

\subsection{The quantum field measurement problem}\label{RelMeasProblem}
Having reviewed the nonrelativistic  measurement problem, we now consider its relativistic generalization. Let us call this the \textit{Quantum Field Measurement Problem}: How are we to understand/model the measurement process which connects the quantum field states in our laboratories to a string of classical data constituting the outcomes of each single-shot measurement? As in the nonrelativistic case, we can distinguish the philosophical and pragmatic aspects of this question, i.e., understanding vs. modeling. However, unlike in the nonrelativistic case, there is at present no consensus on how measurements of quantum fields ought to be modeled mathematically (let alone how they should be understood philosophically).

This is not to say that we know nothing of how measurements of quantum fields ought to be modeled. Indeed, much progress has been made investigating different measurement schemes for quantum fields~\cite{Polo, fewster1, Earman}. It is natural to demand that our measurement theory respects the central `commandments' of relativity: covariance, causality and locality. Such considerations already rule out some of the mathematical tools that we could use to model nonrelativistic quantum measurements. For example, projective measurements in QFT are incompatible\footnote{While philosophical commentary is beyond the scope of this paper, it is worth noting that relativistic considerations can (and likely should) be used as a litmus test for existing answers to the non-quantum field measurement problem. Specifically, we ought to favor those solutions which can be extended to a relativistic setting. By this criteria, solutions to the measurement problem which rely heavily on projective measurements and wavefunction collapse score poorly.} with its relativistic nature:  they are not properly localizable~\cite{Redhead1995} and enable superluminal signaling even in simple setups~\cite{Sorkin,Dowker,Dowker2,borsten}. Moreover, if one persists in using projective measurements in QFT, even as an effective tool, the desired operations are conceptually ill-defined~\cite{alvaro,Adam} and plagued with divergences and other similar practical problems. For all these reasons it has been well established  that we need a new (or at least refined) measurement theory for quantum fields different from that of nonrelativistic quantum mechanics. 

Just as in the nonrelativistic setting, it is helpful here to think of the sequence of interactions which mediate between the measured system (here, a quantum field) and us, the observer. From high-energy physics experiments at particle colliders to the absorption of light at the human retina, quantum fields are subject to measurements where data is extracted through their interaction with localized probes (e.g., atoms being excited by the electromagnetic field, Geiger counters detecting gamma rays, atomic nuclei emitting/absorbing neutrinos via nuclear $\beta$-decay etc). Thus, it is natural to model the first step in this sequence as the quantum field interacting with a localized probe of some sort. This then raises the following questions: How exactly are these localized probes supposed to be modeled? Moreover, how are they to interact with the quantum field? And, perhaps more importantly, to what extent do we need to model this field-probe interaction relativistically?

This last question hits on an important point: Where should we place the \textit{relativistic cut}? Analogously with the Heisenberg cut discussed above, we define the relativistic cut as the point where we can stop modeling the measurement process using quantum field theory. That is, where along the measurement process can we stop modeling the intermediary systems\footnote{E.g., the quantum field being probed, the electrons in an atom, the measurement apparatuses employed to measure the atomic detector, the circuitry and transistors that follow, etc.} as quantum fields? Our understanding of the \textit{relativistic cut} parallels the understanding of the Heisenberg cut laid out in the previous subsection. Namely, we are thinking of the relativistic cut as a nonfundamental theoretical construct. Indeed, we believe the world to be quantum field theoretic through-and-through. Nonetheless, the relativistic cut is to play an essential foundational role in bridging the gap between a QFT description of the world and classically-described experiments and observers. As we claimed above, a physical theory is not fully understood until the observers and their experiments are identified (even if perhaps only in an emergent or approximate regime) within it. 

Suppose that, against the above discussion, we proceed without a relativistic cut, modeling the measurement of a QFT entirely from within a QFT setting. That is, suppose that our measurement process is modeled as quantum fields interacting with quantum fields interacting with quantum fields, etc., until we arrive at a quantum field theoretic description of the experimenter's notepad. This has brought us no closer to being able to extract classical information from this account of our measurement. To do so would be to presume a preexisting solution to the quantum field measurement problem which, of course, we do not yet have. More will be said about this later, but by our reckoning, this is the precise point where the---otherwise impeccable---Fewster-Verch framework~\cite{fewster1,fewster2,fewster3} falls short of fully solving the quantum field measurement problem. The Fewster-Verch approach relies on the fact that, quoting \cite{FewsterRQITalk3}, ``someone, somewhere, knows how to measure something''. That is, their view is that the serious problem of addressing the full measurement problem can be postponed while the links of the measurement chain are analysed individually. During this link-by-link study it is assumed that at some later point along the chain, a measurement outcome is registered. The Fewster-Verch does not seek to explain/model how measurement outcomes are registered (leaving this instead to ``someone, somewhere''). Without a precise description of how this happens in an experiment, in our opinion the Fewster-Verch framework is agnostic about the inevitable relativistic cut.

As the above discussion suggests (at least in the absence of a fundamental quantum field theoretical description of measurement apparatuses and the observers) an explicit and formalized handling of the relativistic cut is essential to any complete solution to the quantum field measurement problem. One may argue that, once we have spelled out exactly how and where we are taking the relativistic cut,  there is practically no more to be said: The measurement theory of QFT is just whatever is induced when our nonrelativistic measurement theory is applied after any late-enough\footnote{Recall our earlier discussion about how the validity of a Heisenberg cuts needs to be evaluated within the context of the whole measurement procedure. There are always ways of continuing an experiment where the quantum systems recohere, thereby invalidating the cut approximation. Conceivably, the same sort of thing can happen with a relativistic cut. In general it is going to be non-trivial to establish when a relativistic cut is considered to be late-enough, but in some setups such as an atomic probe detecting an electormagnetic field excitation it seems reasonable to consider the internal dynamics of the atomic probe as nonrelativistic even though the electromagnetic field is or even if the trajectory of the atom is relativistic.} relativistic cut.

Having argued for the need to place a relativistic cut somewhere to fully understand the measurement of quantum fields, we next ask: where then is it reasonable to place a relativistic cut? As motivated above, in practice, information is extracted from quantum fields via their interaction with localized probes. Examples include: atoms, photo-detectors, Geiger counters, the human retina, etc. To the best of our knowledge, the internal dynamics of these systems are well-modeled nonrelativistically. In that sense, we are justified in applying our nonrelativistic measurement theory to these probes, at least after their interaction with the quantum field. Indeed, nonrelativistic atomic states are exactly the sort of things our nonrelativistic measurement theory has been experimentally validated on. Moreover, recent works have shown that when prescribed carefully, internally\footnote{We remark the word `internally' because the probes can certainly be placed in a rocket going at a large speed with respect to the lab's frame and have relativistic motion in an arbitrary curved spacetime. Still their internal dynamics does not require relativity to yield accurate predictions.} nonrelativistic probes (such as Unruh-Dewitt detectors) can both model realistic experimental setups and respect the covariance and causality of the QFT they couple to~\cite{martin-martinez2015, TalesBruno1, TalesBruno2, PipoMaria}.

Thus in many (but certainly not all) cases it will be reasonable to place our relativistic cut somewhere in the vicinity of the local probe. What remains is to provide an explicit treatment of the cut. We consider the following  two options, one ideal and one more practical. We could: 1)  model the probe as fully relativistic during its interaction with the field, then assign it a nonrelativistic state only after the interaction via some nonrelativistic approximation scheme or 2) model the probe's internal state as nonrelativistic from the start, but design its interaction with the field so as to preserve the causality, locality and covariance of the underlying quantum field theory as much as possible. One can see the first option as what a completion of the  Fewster-Verch framework could achieve, at least in principle. The second option, by contrast, is what the Unruh-DeWitt-like particle detector models already do~\cite{martin-martinez2015, PipoMaria, TalesBruno1, TalesBruno2}. %\tcr{On the first option the relativistic cut is placed after the probe's interaction with the quantum field, but before any of the probe's further interactions. On the second option the relativistic cut is placed in a way ``during'' the probe's interaction with the quantum field.}

Let us give some commentary on the feasibility of the first approach described above. For concreteness, let us particularize momentarily to the case of an atom probing a quantum field. To overview, in this case we would model the quantum field as interacting with a fully relativistic atom (i.e., a collective, non-perturbative, bound state of QED). Following this interaction, we can trace out the probed field's degrees of freedom to find the (still fully relativistic) reduced state of the atom. By some yet-to-be-specified  scheme we could then arrive at a something that is really well modeled by  the (proven accurate enough by more than a century of experiments) nonrelativistic atomic state. Such a state would then be within the well-known domain of atomic experimental physics.

Is this ideal approach feasible? While this approach has notable merits in terms of the preservation of the causality and locality of the underlying quantum field theory, its feasibility is still questionable. As a first issue, bound states of QED are remarkably difficult to either simulate or treat analytically. This is very difficult for bound states in isolation, let alone ones interacting with an external field. Secondly, the process of translating between nonrelativistic atomic states and bound QED states is largely still a mystery. As such the ``nonrelativistic approximation scheme'' suggested above is critically underdeveloped at present.%\footnote{\tcr{One may be tempted at this point to continue with the Fewster-Verch framework but simplify the bound QED state to something more manageable (albeit less realistic). For instance, the Ruep model \cite{ruep2021} identifies the probe system with a local mode of a free theory. As we will discuss later however, this simplification gives a poor model of the local probes used in realistic experiments. This thereby breaks the all-important connection between our measurement theory and real-life experimental procedures.}}.\Dan{New footnote here}

Thus, out of these feasibility concerns, one may be driven to consider the second more practical option. Recall that for this option the internal degrees of freedom of the probe are to be modeled nonrelativistically throughout its interaction with the field. Since quantum field theory itself provides us with no prescription for how relativistic and nonrelativistic systems ought to interact, there is a great deal of freedom in how one might model such an interaction. As a guide, we may rely on a desire to preserve the central concepts of relativity (covariance, causality and locality) as much as possible. Moreover, we may be guided by a desire to match well the light-matter interaction as it is understood in quantum optics~\cite{Pablo}, or even the processes of nuclear $\beta$-decay that can be used to detect neutrinos~\cite{MatsasNeutrinos, perche2021antiparticle}. Under such guidance one is quickly led to the general family of particle detectors that go under the name of Unruh-DeWitt-like models.

\section{A Tale of Two Detector Models}\label{twodetectormodels}
In this section, we review the technical background that is relevant to the comparison between the usual Unruh-DeWitt approach to particle detector models and the Fewster-Verch framework for measurements in quantum field theory.

\subsection{The UDW model}\label{UDWsection}

The Unruh-DeWitt (UDW) model is arguably the most popular tool used to define a measurement framework for quantum fields in the context of relativistic quantum information. The probe (also often referred to as a ``particle detector'') is given by a quantum system strongly localized around some timelike trajectory in spacetime, which defines the region where the detector can couple to the quantum field of interest. In its simplest presentation, the field is given by a minimally coupled real scalar quantum field in $D = n+1$ spacetime dimensions in a globally hyperbolic spacetime, and the full dynamics can be defined by prescribing an action,
\begin{equation}\label{prototypeaction}
    S = S_\phi + S_d + S_I,
\end{equation}
where $S_\phi$ and $S_d$ correspond to the free dynamics of field and detector respectively, and $S_I$ encodes the coupling between them. The free action for the scalar field is taken to be,
\begin{equation}
    S_\phi = -\dfrac{1}{2}\int\dd^Dx\sqrt{-g}\left(\nabla^a\phi\nabla_{a}\phi + m^2\phi^2\right),
\end{equation}
which, upon extremization, leads to the well-known Klein-Gordon equation $(\nabla^a\nabla_{a} - m^2)\phi = 0$. The field is then decomposed into a complete set of modes $\{u_{j}(\mf{x})\}$ constituting solutions to the Klein-Gordon equation as\footnote{Naturally, the symbol $\sumint$ here means that the mode sum written in~\eqref{theQField} is to be understood in a generalized sense: there are situations where one has a discrete basis of modes (when the field is in a cavity with Dirichlet boundary conditions, for example), while in other contexts the label for each mode can be continuous (in a plane-wave decomposition of the field in free space, for instance). In the latter case, sums are replaced by integrals, and Kronecker deltas are replaced by Dirac deltas in expressions such as~\eqref{CCRcreationannihilation} and~\eqref{completesetconditions}.},
 \begin{equation}\label{theQField}
        \hat{\phi}(\mf x) = \sumint_j\left(\hat{a}^\dagger_{j} u_{j}^*(\mathsf x)+\hat{a}_{j}^{\phantom{\dagger}} u_{j}( \mathsf x) \right),
    \end{equation}
where, in the usual process of canonical quantization, we have already replaced  the amplitudes associated to each mode in the general solution by creation and annihilation operators $\hat{a}^\dagger_{j}$, $\hat{a}_{j}$ satisfying canonical commutation relations
    \begin{align}\label{CCRcreationannihilation}
       \big[\hat{a}_{j}^{\phantom{\dagger}},\hat{a}^\dagger_{k}\big] &= \delta_{jk} \openone\nonumber,\\
        \big[\hat{a}_{j}^{\phantom{\dagger}},\hat{a}^{\phantom{\dagger}}_{k}\big] &= 0 ,\\
        \big[\hat{a}_{j}^{{\dagger}},\hat{a}^\dagger_{k}\big] &= 0.\nonumber
    \end{align}
This process is equivalent to imposing the canonical commutation relations between the field and its conjugate momentum at a given Cauchy surface if we also impose that that the basis modes $\{u_{j}(\mf{x})\}$ satisfy the normalization condition
\begin{align}\label{completesetconditions}
    (u_{j}, u_{k})_{\text{K.G.}} =& -(u_{j}^\ast, u_{k}^\ast)_{\text{K.G.}} = \delta_{jk}, \nonumber\\
    (u_{j}, u_{k}^\ast)_{\text{K.G.}} =& (u_{j}^\ast, u_{k})_{\text{K.G.}} = 0.
\end{align}
The Klein-Gordon inner product $(\cdot, \cdot)_{\text{K.G.}}$ is a bilinear form defined for solutions $f_1, f_2$ to the Klein-Gordon equation. It is given by, 
\begin{equation}\label{KGinnerproduct}
    (f_1, f_2)_{\text{K.G.}} = \ii \int_{\Sigma}\dd^n x\sqrt{|h|}n^a\left(f^\ast_1\nabla_{a} f_2 - f_2\nabla_{a} f^\ast_1\right),
\end{equation}
with $\Sigma$ being a Cauchy surface, $n^a$ being its future-pointing unit normal vector, and $\dd^n x\sqrt{|h|}$ being the induced invariant volume element on $\Sigma$.\footnote{By virtue of Stokes' theorem and using the fact that $f_1$ and $f_2$ are solutions to the Klein-Gordon equation, it is easy to see that \eqref{KGinnerproduct} is independent of the choice of Cauchy surface $\Sigma$.} Finally, the vacuum state associated to the choice of modes in Eq.~\eqref{theQField} is defined as the state $\ket{0}$ that is annihilated by all the annihilation operators,
\begin{equation}
    \hat{a}_{j}\ket{0} = 0 \,\,\,\,\,\forall\,\, j,
\end{equation}
and the Fock space of the field is then built by repeatedly acting on the vacuum with creation operators.

The interaction between field and detector in a Unruh-DeWitt-like model can be described by the interaction action,
\begin{equation}\label{prototypeinteraction}
    S_I = -\lambda\int\dd^Dx\sqrt{-g}\Lambda(\mf{x})\mu(\tau)\phi(\mf{x}).
\end{equation}

If one fixes a coordinate frame $(t, \bm{x})$ with $t$ being a timelike coordinate, this action yields the following  \emph{interaction Hamiltonian} (already quantized and in the interaction picture, see~\cite{TalesBruno1} for details),
    \begin{equation}\label{HamiltonianT}
        \hat{H}_{I}^{t}(t) = \lambda \int_{\mathcal{E}_{t}}\!\!\!\dd^n \bm{x} \sqrt{-{g}}\:\Lambda(\mathsf{x})\hat{\mu}(\tau)\hat{\phi}({\mathsf{x}}),
    \end{equation}
    with $\mathcal{E}_t$ denoting the (spacelike) surfaces of constant $t$ in the coordinates $\mf x = (t,\bm x)$. Here, $\lambda$ is a coupling constant, $\Lambda(\mf{x})$ is a spacetime smearing function containing the information about the localization of the detector in space and time, and $\hat{\mu}(\tau)$ is an operator acting on the detector's Hilbert space (i.e., an internal degree of freedom of the probe system), with $\tau$ representing the proper time that parametrizes the trajectory around which the interaction is centered (which we will call the proper time of the detector). 
    
    In the most widely used setup, the detector is modeled as a two-level system---i.e., a qubit---, with $\hat\mu(\tau)$ corresponding to the monopole operator $\hat{\mu}(\tau) = \hat{\sigma}^+(\tau) + \hat{\sigma}^-(\tau)$, and $\hat{\sigma}^\pm$ being the $SU(2)$ ladder operators satisfying the commutation relations $\comm{\hat{\sigma}^+}{\hat{\sigma}^-} = \hat{\sigma}_z$. In that case, the free dynamics generated by the detector's free action, $S_d$, makes the ladder operators evolve according to $\hat{\sigma}^\pm(\tau) = \hat{\sigma}^\pm e^{\pm\ii\Omega\tau}$, with $\Omega$ representing the detector's proper energy gap (i.e., the energy gap between its ground and excited states as measured in its proper frame). This is what is most commonly denoted by the UDW model: a localized two-level system that couples to a real scalar field around a fixed trajectory in spacetime. However, it is also possible to have the detector as being a (bosonic) harmonic oscillator, in which case one can take the operator $\hat{\mu}(\tau)$ as the position operator within the harmonic oscillator's Hilbert space. In this context, for an oscillator with characteristic frequency also given by $\Omega$, one would have \mbox{$\hat{\mu}(\tau) = \hat{a}e^{-\ii\Omega\tau} + \hat{a}^\dagger e^{\ii\Omega\tau}$}, with $\hat{a}^\dagger, \hat{a}$ being the bosonic creation and annihilation operators satisfying the usual canonical commutation relation $\comm{\hat{a}}{\hat{a}^\dagger} = \mathds{1}$. 
    
    For both the qubit and the harmonic oscillator, the Hamiltonian of the form~\eqref{HamiltonianT} corresponds to the generator of time translations in the coordinate $t$ in the interaction picture. This means, in particular, that the time evolution operator generating translations in $t$ \emph{in the interaction picture}\footnote{Recall that, in this picture, operators evolve according to their free dynamics and states evolve with the interaction Hamiltonian.} is given by,
    \begin{equation}\label{timeevolutionoperator}
        \hat{\mathcal{U}}_t = \mathcal{T}_t\exp\left(-\dfrac{\ii}{\hbar}\! \int\!\dd t\, \hat{H}_I^t(t)\right),
    \end{equation}
with $\mathcal{T}_t \exp$ being the time-ordered exponential\footnote{Despite the apparent dependence on a choice of time parameter, for pointlike detectors (and more generally, for interactions that are microcausal---see next footnote) the time evolution operator defined in~\eqref{timeevolutionoperator} is actually frame independent. However, this is not so for spatially smeared UDW detectors. For a deeper discussion on this, see~\cite{TalesBruno2}.} according to the coordinate time $t$.

Since it was first introduced, the UDW model has been widely adopted as a simple yet effective way to make sense of the extraction of local information from a quantum field. Firstly,  (when treated carefully~\cite{Polo}) it yields an operational measurement theory for quantum fields in terms of indirect measurements performed on the probes. Importantly, this measurement theory  does not incur in the same problems that a more naive model---say, based on projective measurements performed directly on field states---would lead to (as pointed out in~\cite{Sorkin}). It has also given a powerful operational means to tackle several important aspects of QFT, such as the observer-dependence of the concept of a particle in QFT (made manifest, for instance, by the Unruh effect) and the exploration of the entanglement structure of the vacuum state of a field theory in possibly curved spacetimes via entanglement harvesting~\cite{Pozas-Kerstjens:2015, Pozas2016, Henderson2018}. 

Besides its appeal as a tool to explore foundational aspects in the interplay between quantum information and quantum field theory, the qubit-based UDW model is also of great practical importance thanks to its ability to describe relevant features of the interaction between atoms and the electromagnetic field~\cite{Richard}. This makes it highly amenable to model experimental setups that can be routinely explored in atomic physics, quantum optics, and superconducting circuits. This is especially the case if 1) we are primarily interested in studying one individual atomic transition (which motivates the simplification to a two-level system) and 2) the phenomenology investigated does not depend on the exchange of angular momentum between atom and field (making the replacement of the electromagnetic field by a real scalar field a fair approximation). 

Given its prominent role as a testbed for concrete investigations of foundational concepts in quantum field theory, it is of utmost importance to evaluate whether  the UDW model is compatible with the requirements that a relativistic measurement theory must satisfy. These include notions such as locality, causality and general covariance, as we discussed in Sec.~\ref{RelMeasProblem}. In the case of pointlike UDW detectors---where the spacetime smearing $\Lambda(x)$ becomes infinitely localized at a single point for each fixed value of the detector's proper time---it is known that the model is fully causal and covariant~\cite{TalesBruno2,martin-martinez2015,PipoMaria}. This can be traced back to the fact that, for the case of pointlike detectors, the interaction is local (the detector's degree of freedom couples to the field at only one point at each instant of time) and microcausal\footnote{That is, the interaction Hamiltonian density commutes with itself at spacelike-separated points~\cite{TalesBruno2}.}. 

Furthermore, for a smeared UDW detector where one has a nontrivial spatial profile for the probe at each instant of time, it is also possible to show that the coupling between field and detector can be made covariant~\cite{TalesBruno1}. On the other hand, the simplifying assumption that the smeared detector is still described by just a single degree of freedom now coupling to the field in a spatially extended region technically leads to a small degree of nonlocality. This is manifest through a violation of microcausality of the Hamiltonian density within the light-crossing time in the detector. In turn, this has been demonstrated to make the predictions for the dynamics as computed with a smeared UDW model only \emph{approximately} covariant. This is made explicit by showing that, when the interaction is not microcausal, time evolution operators computed between the same past and future Cauchy surfaces but using different time coordinates (which one would expect to be a purely calculational artefact with no physical impact on the actual dynamics) can lead to different physical predictions for the model. The degree of breaking of covariance in the model can be made precise in terms of the field's Wightman function and the detector's smearing~\cite{TalesBruno2}.

Fortunately, however, this breaking of covariance is not as dramatic as it might sound at first: the scales where one would start seeing a relevant impact of this breaking of covariance from the smeared UDW model correspond to the same scales where one should already expect the approximations involved in the model (the simplification of a single quantum degree of freedom extended over some region) to not be reliable in the first place. Of course, it is true that no Unruh-DeWitt-like model should be portrayed as a first-principle description of quantum field theory at a more fundamental level. Their applicability is restricted, for instance, by the very requirement of an appropriate effective theory of the detector system in terms of a single, localized degree of freedom. The energy scales at which the Unruh-DeWitt model is a good effective theory for a measurement apparatus in QFT are very well understood~\cite{Pablo, TalesBruno2, PipoMaria} and they lay within the realm of measurements carried out even in our highest-energy physics experiments.

At this point, we want to draw our attention to a relevant claim made in~\cite{ruep2021}: that Unruh-DeWitt-like detector models are either \emph{singular} or \emph{nonlocal}, and are thus of limited applicability when dealing with foundational questions in QFT. The rationale behind this claim is based on the behavior of the underlying classical equations of motion of the dynamical system described by an action like~\eqref{prototypeaction} with an interaction given by~\eqref{prototypeinteraction}. If the UDW detector is pointlike, the interaction behaves as a source term for the field that diverges at a point for each value of the detector's proper time, thus making the equation singular; if, on the other hand, the detector is smeared, the interaction now couples a single degree of freedom of the detector to the field evaluated at spacelike-separated points simultaneously, which renders the equation of motion nonlocal. However, we believe the singular/nonlocal nature of the classical equations of motion of the UDW detector model is not central to the validity of the conclusions that can be drawn from it, for the following reasons.

First, the non-locality problem has been partially addressed in the earlier paragraphs. We reiterate that even for smeared detectors, the degree to which the non-locality impacts the predictions of the theory is suppressed precisely in the most common regimes where one would normally use such detector models---in particular, when the predictions are taken in time scales longer than the detector's light-crossing time. 

Second, regarding singularity, in the way the calculations are routinely performed, one works with a time evolution operator such as~\eqref{timeevolutionoperator} in the interaction picture, where operators evolve according to the \emph{free} equations of motion. Therefore, at least at a perturbative level, it is not at all clear where the singular nature of the fully coupled equations of motion plays a role in the conclusions from the model. Even if the field's equations of motion itself are singular, both the \emph{response} of the detector and the field state perturbation are regular, and thus show no pathologies or divergences even when the interaction is supported just along a single timelike worldline, as long as one uses smooth switching functions in the detector's proper time\footnote{In some cases this may include a temporary regularization by a spatial profile, after which one takes the limit of vanishing size for the detector~\cite{Jorma}.}~\cite{Jorma,Satz2007}. 

%As we shall see in more detail later, one can give another simple variant of the two-level UDW detector, where the quantum system describing the probe is taken to be a harmonic oscillator instead of a two-level system. For obvious reasons, we shall refer to this model as the harmonic-oscillator UDW detector. The reasons why one may be motivated to do this will be explored in more depth in the following sections.

\subsection{Fewster-Verch framework and the Ruep model}

The most rigorous axiomatic treatment of quantum field theory is found in the framework of algebraic quantum field theory (AQFT). It is therefore reasonable to expect that a first-principle-based formalism for a measurement theory for quantum fields should be phrased in this language. The general Fewster-Verch framework~\cite{fewster1} offers a very promising approach to this problem. This framework has indeed pointed towards a consistent resolution to a long-standing conceptual problem regarding the notion of a state-update rule for measurements in quantum fields that does not incur in causality violations from superluminal signalling~\cite{fewster3}. More recently, it has also been used~\cite{ruep2021} for an investigation on entanglement harvesting, using an explicit model for the probe theory consisting of localized modes of a free scalar field. Above, we referred to this particular model of local probes as the \emph{Ruep model}. In what follows, we will attempt to give a self-contained review of the relevant concepts involved in the formulation of the framework, and how it is realized concretely in the Ruep model.

In AQFT, the focus is placed first and foremost on the \emph{local algebra of observables}, which can be understood intuitively as the collection of field degrees of freedom that can be assigned to local subregions of spacetime. We then make further requirements on the algebra based on information about the \emph{causal structure} of the underlying spacetime, in order for it to describe a relativistic QFT. It is therefore important that we fix some jargon on causal structure in Lorentzian spacetimes, to which we now turn.

We will closely follow the terminology used in~\cite{fewster1}. Let $\mathcal{M}$ denote a spacetime described by a differentiable manifold with a Lorentzian metric admitting a choice of time orientation. A curve in $\mathcal{M}$ is said to be \emph{causal} if its tangent vector is timelike or null at every point. Given a point $x \in \mathcal{M}$, the \emph{causal future/past} of $x$ is the set of all points that can be reached from $x$ by a future/past pointing causal curve (including $x$ itself), and is denoted by $J^{+/-}(x)$. For a subset $S \subseteq \mathcal{M}$, we denote \mbox{$J^{+/-}(S) = \bigcup_{x\in S}J^{+/-}(x)$}, and also \mbox{$J(S) = J^{+}(S)\bigcup J^-(S)$}. The \emph{causal hull} of a subset $S$ is written as $\text{ch}(S)$, and is defined by \mbox{$\text{ch}(S) \coloneqq J^+(S)\bigcap J^-(S)$}.  The \emph{causal complement} of a subset $S\subseteq \mathcal{M}$ is defined as \mbox{$S^\perp = \mathcal{M}\setminus J(S)$}, and only contains points which are spacelike-separated from all points in $S$. 

A set is said to be \emph{causally convex} if it is equal to its causal hull, which in particular means that it contains all causal curves that both begin and end within the set. Two subsets $S, T \subseteq \mathcal{M}$ are said to be \emph{causally disjoint} if there is no causal curve going from one set to another, which means that $S \subset T^\perp$ (or, equivalently, $T \subset S^\perp$). The \emph{domain of dependence} of a set $S$, denoted by $D(S)$, consists of all points $x \in \mathcal{M}$ such that \emph{every}\footnote{Note that while $J(S)$ denotes the set of points that can receive (or send) \emph{some} influence---here understood as signals sent through causal curves--- from (or to) events in $S$, $D(S)$ is composed only of points $x$ such that \emph{all} influence reaching $x$ can be traced back to some point in $S$.} inextendible causal curve containing $x$ crosses $S$. An \emph{achronal surface} is a codimension $1$ submanifold $A \subseteq \mathcal{M}$ such that, for any two points $x, y \in A$, there is no timelike curve connecting $x$ to $y$. A \emph{Cauchy surface} is an achronal surface $\Sigma$ such that its domain of dependence is the entire spacetime: $D(\Sigma) = \mathcal{M}$. If a spacetime admits a Cauchy surface, it is called \emph{globally hyperbolic}. The requirement of global hyperbolicity is important because it allows for a well-posed initial-value formulation of the dynamics of fields on $\mathcal{M}$, where the classical evolution throughout the whole $\mathcal{M}$ is entirely determined by initial data defined on the vicinity of a Cauchy surface. 

The starting point to construct a quantum field theory in a globally hyperbolic (possibly curved) spacetime $\mathcal{M}$ is to define a unital $\star$-algebra\footnote{An \emph{algebra} is a vector space equipped with a product. A \mbox{\emph{$\star$-algebra}} is an associative algebra over the complex numbers, equipped with an operation (the $\star$-operation) with the properties of the adjoint. A \emph{unital} algebra is an algebra equipped with an identity element.} $\mathcal{A}(\mathcal{M})$, along with a collection of sub-$\star$-algebras $\mathcal{A}(\mathcal{M}; N)$ for every causally convex subset $N\subset\mathcal{M}$, with the same unit element as  $\mathcal{A}(\mathcal{M})$. Any local relativistic QFT is then expected to comply with a few basic properties: 

\begin{enumerate}
    \item \emph{Isotony}: if $N_1 \subseteq N_2 \Rightarrow \mathcal{A}(\mathcal{M}; N_1) \subseteq \mathcal{A}(\mathcal{M}; N_2)$. 
    \item \emph{Compatibility}: If $N$ is an open causally convex subset of $\mathcal{M}$, and $\mathcal{A}(\mathcal{M})$ is defined, then $\mathcal{A}(N)$ is also defined, and there is a unit-preserving algebraic homomorphism $\alpha_{\mathcal{M}, N}:\mathcal{A}(N) \rightarrow \mathcal{A}(\mathcal{M})$ whose image coincides with $\mathcal{A}(\mathcal{M}; N)$. In simple terms, if  one thinks of $N \subset \mathcal{M}$ as a manifold in its own right, the QFT defined within it must be equivalent to the restriction of the QFT in $\mathcal{M}$ to the region $N$.
    \item \emph{Time-slice property}: $\mathcal{A}(\mathcal{M}; N) = \mathcal{A}(\mathcal{M}; N')$ whenever $N$ is contained in the domain of dependence of $N'$. This axiom communicates the existence of dynamics with a well-defined initial-value formulation, in such a way that, in particular, the algebra on the whole spacetime is fully determined by the algebra on an open set containing a Cauchy surface.
    \item \emph{Einstein causality}: if $N_1$ and $N_2$ are causally disjoint, all elements of $\mathcal{A}(\mathcal{M}; N_1)$ commute with all elements of $\mathcal{A}(\mathcal{M}; N_2)$. This is essentially the axiom of \emph{microcausality}, which forbids superluminal signalling by imposing that spacelike-separated elements of the algebra must commute. 
    \item \emph{Haag property}: let $K$ be a compact subset of $\mathcal{M}$. Let $A \in \mathcal{A}(\mathcal{M})$ be such that $A$ commutes with every element of $\mathcal{A}(\mathcal{M}; N)$ whenever $N \subset K^\perp$. Then, the Haag property postulates that $A \in \mathcal{A}(\mathcal{M}; L)$ whenever $L$ is a connected open causally convex region containing $K$. Roughly, this says that there is no way for $A$ to commute with everything in $K$ except by $A$ being localized in a region space-like to $K$.
\end{enumerate}
Finally, in the algebraic approach, states are seen (rather abstractly) as maps from elements of the algebra to complex numbers\footnote{In usual quantum mechanics, one would normally refer to ``observables'' more restrictively as the set of \emph{self-adjoint} elements of the local algebra of observables, in which case all expectation values are real. For generality, however---to allow for a theory to have a complex scalar field as an element of the algebra, for example---we include the possibility that the output of $\omega$ is complex.} \mbox{$\omega: \mathcal{A}(\mathcal{M})\rightarrow\mathds{C}$}, which has the interpretation of taking observables on the algebra and sending them to their expectation values. Such maps are required to be positive ($\omega(A^\star A)\geq 0$ for all $A \in \mathcal{A}(\mathcal{M})$) and normalized ($\omega(\mathds{1}) = 1$, with $\mathds{1}$ being the identity element of the algebra). 

With that all said, one can now formulate local measurements on quantum fields in the Fewster-Verch framework. In this framework, one will generically have a \emph{target system} theory\footnote{As we are using the term, a ``theory'' consists of the combination of the local algebra of observables together with the prescription of states (e.g., a vacuum). However, for the ease of notation, from now on we make a slight abuse of language by referring to the theory with the same notation as we would use for just the algebra.} $\mathcal{S}(\mathcal{M})$ describing the target field we want to extract information from, a \emph{probe} theory $\mathcal{P}(\mathcal{M})$ describing the physical system modeling the measurement device, and a \emph{coupled} theory $\mathcal{C}(\mathcal{M})$ encoding the effect of the interaction between the target and the probe. All theories $\mathcal{S}(\mathcal{M})$, $\mathcal{P}(\mathcal{M})$, $\mathcal{C}(\mathcal{M})$ are described by local relativistic QFTs through their local algebra of observables as in the framework outlined above. The coupling between target and probe is localizable to a compact \emph{coupling zone} $K$ where the interaction between probe and target is turned on. The coupled theory $\mathcal{C}(\mathcal{M})$ is---as the name suggests---a coupled version of the simple uncoupled juxtaposition of target and probe theories \mbox{$\mathcal{S}(\mathcal{M})\otimes\mathcal{P}(\mathcal{M})$}. This notion is implemented by postulating the existence of an isomorphism from $\mathcal{C}(\mathcal{M}; L)$ to $\mathcal{S}(\mathcal{M}; L)\otimes\mathcal{P}(\mathcal{M}; L)$ for any region $L$ such that $L \bigcap \text{ch}(K) = \emptyset$. The existence of such isomorphism allows one to also define an \emph{auto}morphism from the uncoupled theory to itself given by a \emph{scattering map} $\Theta: \mathcal{S}\otimes\mathcal{P}\rightarrow\mathcal{S}\otimes\mathcal{P}$, which should be understood as implementing the effect of the interaction on an input defined on an in-region $\mathcal{M}\setminus J^+(K)$, and returning the output on an out-region $\mathcal{M}\setminus J^-(K)$. If we then start the target and probe in a product state $\omega_{\mathcal{S}}\otimes\sigma_{\mathcal{P}}$, the scattering map $\Theta$ provides a way to define an ``updated state'' $\sigma'_{\mathcal{P}}$ for the probe given implicitly by,
\begin{equation}\label{updatedstate}
    \forall P\in\mathcal{P}(\mathcal{M}),\ \ \sigma'_{\mathcal{P}}(P) = (\omega_{\mathcal{S}}\otimes\sigma_{\mathcal{P}})(\Theta(\mathds{1}_{\mathcal{S}}\otimes P)),
\end{equation}
where we have demanded that the equation above be true for all elements $P$ of the local algebra of observables of the probe theory $\mathcal{P}$, and $\mathds{1}_{\mathcal{S}}$ here is the unit element of the target system theory $\mathcal{S}$. One could think of~\eqref{updatedstate} as describing the final state of the probe after it went through the interaction with the target system. This type of language could however raise questions as to where/when such change in the states of the probe and system precisely occur, which was part of the kind of concern that led to the issues with ``impossible measurements on quantum fields''~\cite{Sorkin}. It is therefore important to emphasize that a key conceptual point on the formulation of the FV framework is the understanding that the measurement actually occurs in the \emph{coupled} theory. Statements made in terms of the uncoupled theory and its components (target system and probe) separately, as well as the notion of state update of probe and system after the interaction, serve the purpose of describing the process in a language that is amenable to our intuitive picture of a measurement, but should not be taken too literally.

\begin{figure}
\includegraphics[width=0.45\textwidth]{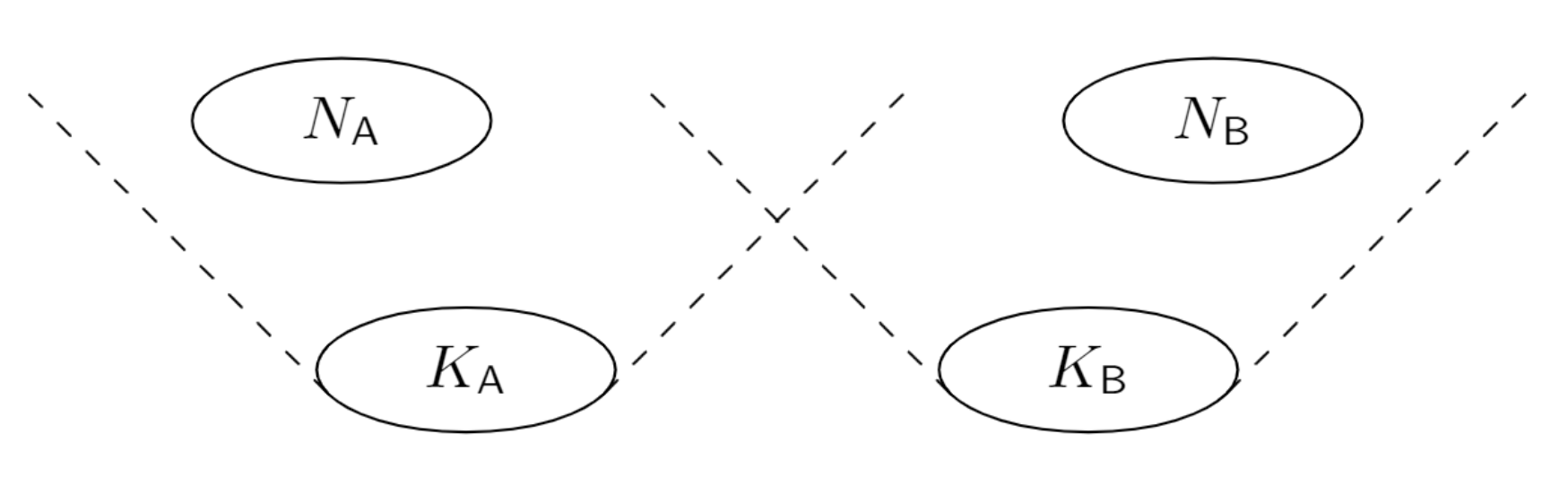}
\caption{Schematic view of a possible configuration of the coupling zones $K_\mathsf{A}$, $K_{\mathsf{B}}$ and processing regions $N_{\mathsf{A}}$ and $N_{\mathsf{B}}$ in a spacetime diagram.}\label{spacetimediagram}
\end{figure}

For the case of entanglement harvesting in particular (which is the main goal of~\cite{ruep2021}), the probe theory is composed of two independent probe fields $\mathcal{P}_{\mathsf{A}}\otimes\mathcal{P}_{\mathsf{B}}$, and the coupling zone splits into two regions $K_{\mathsf{A}}$ and $K_{\mathsf{B}}$ where each of the probes couple to the system of interest separately. The probe fields are initialized in a separable state $\sigma_{\mathsf{AB}} = \sigma_{\mathsf{A}}\otimes\sigma_{\mathsf{B}}$, with the initial state of the target being given by $\omega_{\mathcal{S}}$; the full system is therefore initialized in the state $\omega_{\mathcal{S}}\otimes\sigma_{\mathsf{A}}\otimes\sigma_{\mathsf{B}}$, and is then sent through the scattering map $\Theta$, giving rise to an updated state $\sigma_{\mathsf{AB}}'$ of the uncoupled probe theory $\mathcal{P}_{\mathsf{A}}\otimes \mathcal{P}_{\mathsf{B}}$ defined through Eq.~\eqref{updatedstate}. Finally, agents Alice and Bob, with access to the probe fields $\mathsf{A}$ and $\mathsf{B}$ in \emph{processing} regions\footnote{In the causal future of $K_{\mathsf{A}}$ and $K_{\mathsf{B}}$, respectively.} $N_{\mathsf{A}}, N_{\mathsf{B}}\subseteq \mathcal{M}$ respectively (See Fig.~\ref{spacetimediagram}), will test the entanglement on the joint updated state $\sigma'_{\mathsf{AB}}$ of the two probes by measuring the statistics of the correlations of probe field observables located in their respective processing regions. 

 At this point there is a conspicuous gap in the Fewster-Verch framework: how are the probe fields themselves to be measured? By coupling the quantum field of interest to a probe system that is also described by a quantum field, one is transferring the problem of measurements in QFT from one field (the target system) to another (the probe)---in other words, in the terminology introduced in Sec.~\ref{RelMeasProblem}, one has simply delayed the relativistic cut. Ultimately, without a formal handling of the relativistic cut, the Fewster-Verch framework is not by itself capable of describing how the probes used in the measurement process would, even in principle, generate classical information that could be read by an experimentalist in the processing region.

It is argued in~\cite{ruep2021} that the above picture of entanglement harvesting via local probes is ``model-independent''  because there are as-of-yet no assumptions about the nature of the probed quantum field other than it is a relativistic QFT. However, in order to give concrete predictions, one must provide a more explicit model for a measurement setup (including a model of the probe theory) which follows the general requirements described above and gives us some sense in which an experimenter will extract a concrete value for a measurement outcome from the (QFT) probe.

Ideally, one would wish to model a probe by some localized state of an interacting field theory that contains structures such as bound states or condensates. There is a considerable body of literature in AQFT that deals with several features of interacting theories, with examples that include perturbative approaches~\cite{KasiaPAQFT, KasiaBook}, conformal field theories and integrable models~\cite{CFT, IntegrableModels}, as well as rigorous formulations not just for the free scalar field, but also for linearized gravity~\cite{FewsterLinGrav} and Dirac, Maxwell, and Proca fields~\cite{DiracField, MaxwellField, ProcaAndMore}, among many others. Unfortunately, however, as commented on Sec.~\ref{RelMeasProblem}, handling the emergence of bound systems that may have a nonrelativistic limit from interacting quantum field theories is still a daunting task. Therefore, as a working hypothesis and for the sake of simplicity, in~\cite{ruep2021} a decision is made to model the probes through a real scalar field $\hat{\varphi}$. The quantum theory of the real scalar field is constructed from the algebra of \emph{smeared} field operators $\hat{\varphi}(f)$, with $f\in C_c^\infty(\mathcal{M})$ being a smooth, compactly supported test function. The quantum field $\hat{\varphi}$ consists of an operator-valued distribution that maps test functions $f$ to operators $\hat{\varphi}(f)$, such that for all test functions $f$, $g\in C_c^\infty(\mathcal{M})$, the following properties are satisfied:
\begin{enumerate}
    \item The map $f\mapsto\hat{\varphi}(f)$ is linear over the complex numbers. 
    \item $\hat{\varphi}(f)^\dagger = \hat{\varphi}(f^\ast)$.  
    \item $\hat{\varphi}((\nabla_{a}\nabla^a - m^2)f) = 0$. 
    \item $\comm{\hat{\varphi}(f)}{\hat{\varphi}(g)} = \ii E(f, g)\hat{\mathds{1}}$,
    where the object $E(f, g)$ is defined by,
\begin{equation}
    E(f, g) = \int \dd V\dd V' f(\mf{x}) E(\mf{x}, \mf{x}')g(\mf{x}'),
\end{equation}
with $\dd V = \dd^D x\sqrt{-g}$, and $E(\mf{x}, \mf{x}')$ is the integral kernel that gives the causal propagator (the difference between advanced and retarded Green functions) to the Klein-Gordon equation.
\end{enumerate}
It is useful to think of the smeared field operators as being the effect of integrating the field $\hat{\varphi}(\mf{x})$ at each point in spacetime against a compactly supported smearing function $f$. Requirement number one is simply due to the conceptualization of $\hat{\varphi}$ as a distribution, which by definition means that it is a linear functional on the space of test functions. The second requirement encodes the information that the field is ``real'' (or Hermitian). The third means that $\hat{\varphi}$ satisfies the Klein-Gordon equation as a distribution. Finally, the fourth requirement represents a more covariant version of the ``equal-time'' canonical commutation relations between field and its conjugate momentum on a given Cauchy surface of one's choice. 

To complete the definition of the theory, one now has to prescribe a state---which, as we mentioned earlier, essentially consists of a functional that takes expectation values of the elements of the local algebra. Since our algebra is composed of the unit element and products of smeared fields, a state will be completely defined by the collection of $n$-point functions, 
\begin{equation}
    W_n(f_1, \cdots, f_n) \coloneqq \omega\left(\hat{\varphi}(f_1)\cdots\hat{\varphi}(f_n)\right).
\end{equation}
The simplest possible class of states is given by those such that all $n$-point functions are derived from the $2$-point function\footnote{One could be a bit more general and allow for a nonvanishing $1$-point function as well, so that higher-order $n$-point functions are determined in terms of $1$- and $2$-point functions. However, one can always make the $1$-point function vanish by a simple local field redefinition. Since such field redefinitions do not have any impact on the mixedness or entanglement measures that will be our main interest in the later discussion, we will from now on take the $1$-point function to vanish.} (which, with a slight abuse of language, we will call \emph{Wightman} function). These are known as \emph{Gaussian} or \emph{quasi-free} states. For this class of states, one can recursively compute higher-order $n$-point functions from the Wightman function by building the formal series expansion,
\begin{equation}\label{wicktheorem}
    \sum_{n=0}^\infty \dfrac{\ii^n}{n!}z^n\omega\left(\hat{\varphi}(f)^n\right) = e^{-\frac{z^2}{4}\Delta(f, f)},
\end{equation}
and systematically matching the right and left hand sides of the equation above at each order in the parameter $z$. Here, $\Delta$ is a bilinear symmetric functional on test functions, which can be shown to correspond to the expectation value of the \emph{anti}-commutator of the field:
\begin{equation}
    \Delta(f, g) = \omega(\{\hat{\varphi}(f), \hat{\varphi}(g)\}).
\end{equation}
The expectation value of the commutator is state-independent, due to the covariant canonical commutation relations that are part of the definition of the algebra. 

For quasi-free states, the requirement of positivity\footnote{Recall, $\omega(A^\star A)\geq 0$ for all $A \in \mathcal{A}(\mathcal{M})$.} holds if the bilinear functional $\Delta$ satisfies,
\begin{equation}
    \abs{E(f, g)}^2\leq \Delta(f, f)\Delta(g, g),
\end{equation}
for all real test functions $f$ and $g$.  Conversely, any choice of $\Delta$ under the constraint above leads to a definition of a quasi-free state: its Wightman function is given by, 
\begin{equation}\label{wightmanquasifree}
    W(f, g) = \dfrac{1}{2}\Delta(f, g) + \dfrac{\ii}{2}E(f, g).
\end{equation}
All the odd higher-order $n$-point functions vanish and the even ones are computed from~\eqref{wightmanquasifree} using~\eqref{wicktheorem}.

It is straightforward to see that any vacuum state $\ket{0}$ defined via canonical quantization as explained in Sec.~\ref{UDWsection} is quasi-free. In fact, the converse is also true: namely, any pure quasi-free state with vanishing $1$-point function can be seen as the vacuum state in some Fock-space representation of the algebra of smeared fields. In other words: for any pure quasi-free state $\omega$, one can find a set of mode functions $u_k(\mathsf{x})$ satisfying~\eqref{completesetconditions} and construct a Fock space with a vacuum $\ket{\Omega_\omega}$ such that,
\begin{equation}
    \omega(\hat{\varphi}(\mathsf{x})\hat{\varphi}(\mathsf{x}')) = \bra{\Omega_\omega}\hat{\varphi}(\mathsf{x})\hat{\varphi}(\mathsf{x}')\ket{\Omega_\omega} = \sumint_j u_j(\mathsf{x})u_j^\ast(\mathsf{x}').
\end{equation}
Furthermore, in this Fock space, one can define creation and annihilation operators associated to each mode satisfying the canonical commutation relations~\eqref{CCRcreationannihilation}, which allow the field $\hat{\varphi}$ to be explicitly decomposed as in~\eqref{theQField}. The recipe that allows us to represent the field algebra acting on some Fock space with vacuum state $\ket{\Omega_\omega}$ identified with the quasi-free state $\omega$ of one's choice is known as the \emph{GNS representation}~\cite{GNS, Haag, FewsterRejzner}.

Once the theory (with the algebra of smeared fields and the choice of quasi-free state $\omega$) is given, one can define a local mode of the field via quadrature operators given by
\begin{align}
    \hat{Q} = \hat{\varphi}(f_1), \\
    \hat{P} = \hat{\varphi}(f_2),
\end{align}
where the functions $f_1$ and $f_2$ are compactly supported inside the coupling region $K$ of interest, and we can choose them such that $E(f_1, f_2) = 1$. The commutation relation guarantees that these quadrature operators satisfy,
\begin{equation}\label{LocalMode}
    \comm{\hat{Q}}{\hat{P}} = \ii\mathds{1},
\end{equation}
and thus can be used to single out a single-mode harmonic oscillator from the probe field. Finally, the state of the probe is chosen to be the restriction of the quasi-free state $\omega$ of the probe field to the region $K$ where the interaction takes place.

By using local modes of a free scalar field and restricting the quasi-free state $\omega$ to a local subregion, one can show that the states of the probes must be mixed. This is due to a very general result encoded in the Reeh-Schlieder theorem~\cite{Schlieder, Schlieder2}, which essentially guarantees that any quasi-free state in a QFT will exhibit entanglement between spacelike-separated regions, and therefore its restriction to any local subregion will result in a mixed state. Such mixedness then hinders the ability of the probes to perturbatively harvest entanglement from the field of interest, and it is shown in~\cite{ruep2021} that there must be a finite threshold for the coupling constant between probes and field below which no entanglement is harvested.

We emphasize that the main conclusion obtained in~\cite{ruep2021} on entanglement harvesting---namely, the existence of a finite threshold on the coupling constant below which no entanglement is harvested perturbatively---ultimately boils down to the fact that the initial states for the probes are mixed. It is therefore natural to wonder (as pointed out by the author of~\cite{ruep2021} himself) if a similar effect is also present in the more usual UDW-like detector models if we initialize the probes in mixed states, instead of the more standard pure states (commonly taken as the ground states of the probes). Another related question that is left open in~\cite{ruep2021} is precisely how much mixedness should be expected in a setup with realistic probes, and how large the threshold coupling strength is as a function of the probe mixedness. 

In the following sections, we address these issues on two fronts. Firstly, in Sec.~\ref{RealisticMixedness} we will put into context the claim in~\cite{ruep2021} that the reduced state of any local mode will be mixed. In particular, we will show that given an arbitrary spacetime region and a quantum field of arbitrary temperature we can find local modes supported in that region whose reduced state is arbitrarily pure. Thus, while it is true that local modes are necessarily mixed, there is no non-trivial lower bound on the mixedness in terms of either the regions size nor the ambient field temperature. This arbitrary purity is achieved solely by varying the ``shape'' of the local mode, in a way that will be made precise in the next section. Moreover, we investigate for a local mode with a fixed shape the effects of the mode's size and the ambient temperature on the mode's purity. As we will discuss, the dependence of the local mode's purity on its shape, size, and the ambient temperature all indicate that local field modes of a free theory are not good models of the localized probes which are experimentally available (e.g., atomic probes or other similar probes typically modeled with UDW-like interactions). 

In \cite{ruep2021} this problem is addressed by commenting on the possibility that the probe field $\hat{\varphi}$ is not free, but is subject to an external potential that can act as a spacetime-dependent mass. By making the field very ``heavy'' outside of some localized region, one could then mimic the effect of a cavity implementing approximate Dirichlet boundary conditions on the field; in this setup, the Reeh-Schlieder property of quasi-free states would technically still hold, and thus so would the existence of some bound on entanglement harvesting. We know, however, that in the limit of a perfectly reflecting cavity with Dirichlet boundary conditions on the field inside, the field can effectively be quantized in a product state where we have a vacuum within the cavity, fully decoupled from what happens outside of it, at least for some long enough times~\cite{HighQfactor, HighQfactor2}. We therefore expect---as it was also acknowledged in~\cite{ruep2021}---that the effect of such ``approximate'' cavity model will be to systematically decrease the threshold of mixedness at which entanglement harvesting is possible, to the point where it also eventually becomes negligible.

Secondly, going back to the Unruh-DeWitt detector model, in Sec.~\ref{UDWperturbative} we study how mixedness inhibits entanglement harvesting. We show how one can perturbatively compute the mixedness threshold beyond which no entanglement is harvested---or, conversely, the minimum coupling strength necessary to overcome an initial amount of mixedness. As we will see in further detail, at lowest order, the effect of initial mixedness (which we model by assigning thermal states with finite temperature to the probes) is to simply correct the populations of the first excited states of the detectors. This fact leads to a very straightforward modification of the well-known formulas for entanglement harvesting with initially pure probes, both for two-level detectors and harmonic oscillators. We can then evaluate quantitatively how the mixedness of the probes impacts the amount of entanglement harvested, which readily allows us to identify what is setting the scale for the maximum mixedness (or alternatively, the minimum coupling strength) that can lead to entanglement harvesting perturbatively. Contrary to what one might have expected based on claims from~\cite{ruep2021}, we will see that realistic detectors, with typical levels of mixedness and coupling strengths, are well within the regime where entanglement \emph{can} be harvested perturbatively. This is closely related to the fact that, for typical experimental setups (described, for instance, by atoms or trapped ions used in atomic physics and quantum optics), the state of the probes are, to a very good approximation, given by pure states. It is true that, at a fundamental level, one can never reach perfect purity, since one can never cool down a system to absolute zero temperature in finite time. However, the level of purity achievable in setups in quantum optics is such that the mixedness from the Reeh-Schlieder property plays little to no role in such settings. 

\section{Estimating Realistic Levels of Probe Mixedness}\label{RealisticMixedness}
As discussed in the previous section, one of the central ideas supporting the conclusions of \cite{ruep2021} is the fact that quantum fields (in particular, Reeh-Schlieder states) generically have entanglement between space-like separated compact regions. A direct consequence of this entanglement is that when we consider the reduced field state associated with any compact region, we will find that it is a mixed state. When the global field state is the vacuum, the mixedness of these regions is well understood quantitatively through area laws. What is important, however, for the claims of \cite{ruep2021} is not the mixedness of the field in a compact region, but the mixedness of the local modes of the probe field supported within that region. In \cite{ruep2021} it is shown that the reduced state of any local mode will be mixed (their Lemma 15). That is, if the field is in a Reeh-Schlieder state then every compactly localized mode of a quantum field is in a mixed state. It is this unavoidable mixedness which supports their further claims about the impossibility of perturbative entanglement harvesting for weak enough coupling strength.

However, \cite{ruep2021} provides no quantitative analysis of how mixed local modes must be due to the Reeh-Schlieder property. A quantitative analysis of this mixedness is important because it factors directly into the magnitude of coupling strength needed for entanglement harvesting: if local modes can be arbitrarily pure, then we can potentially extract entanglement using arbitrarily small coupling strengths. This section will provide a quantitative discussion of mixedness due to Reeh-Schlieder-like localization effects. Moreover, we will compare this to the level of mixedness which one can expect due to the ambient temperature of the field to assess how relevant the fundamental \textit{Reeh-Schielder mixedness} is in a realistic scenario.

This quantitative analysis will be carried out in two stages. Firstly, we will search for the minimally mixed local mode compactly supported in some region and given some ambient temperature. As we will see, we can find local modes which are arbitrarily pure in every case. Secondly, we will investigate the balance between mixedness due to localization and mixedness due to temperature, for a number of fixed local mode profiles. As we will see, in practical scenarios, mixedness due to temperature dominates over mixedness due to localization. This is, of course, what one would expect from experimental physics. Temperature is often the limiting factor in producing pure initial states for quantum computing and quantum information processing. 

This section will conclude with some comments comparing the levels of mixedness that one can expect from local modes in a free theory, with the typical mixedness of more realistic local probes constructed from QED (namely, atomic probes or other similar probes typically modeled with UDW-like interactions).

%If our local probe is to be identified with one of these compactly localized degrees of freedom, then it initial state will unavoidable be mixed. \footnote{More generally, one could consider a local probe whose degrees of freedom are built up from several of the field's compactly localized degrees of freedom. A priori, it is thinkable that even if each localized degree of freedom is mixed, the joint state of some collection of them might not be. This is certainly allowed for within the formalism of quantum mechanics; Joint states are often pure when the reduced states are not. However, while this possibility is not considered in \cite{}, it easy to see that their claims extend to this case as well. Proof: Consider any two mode Gaussian state, with the reduced states mixed and the joint state pure. In any such case, we can decompose the joint state into two new modes which are uncorrelated. This process is linear and just amounts to symplectic diagonalization of the Gaussian state's covariance matrix. Since the joint state is pure, each of these new modes must be pure as well. Moreover, since these new modes are a linear combination of the original modes, they will be compactly localized if the original modes were. Thus the existence of multi-mode local pure states imply the existence of single-mode local pure states. But as was proven in \cite{} there are no single-mode local pure states. Thus, the joint state of every collection of localized modes is mixed.}

\subsection{Minimally mixed local field modes}
In this subsection we are interested in identifying the minimum mixedness that a field mode can have while being supported over some compact spacetime region, $\mathsf{x}\in R\subset\mathcal{M}$. In particular, we are interested in how the mixedness of this mode varies as we change both the size of the region and the ambient temperature, $\beta$, of the field. We will argue that for any region and any temperature there are local field modes with non-zero but arbitrarily low mixedness, i.e., arbitrarily high purity. 

%While we take this claim to be true dimension-independently, for simplicity we will consider only the 1+1 dimensional case in full technical detail. Our proof in the 1+1 dimensional case will be given in the next subsection. Following this we will comment on how this proof might be adapted to higher dimensional cases. We believe that our the results can likely be generalized to higher dimensional settings quite easily. As such, in this section we will state the problem in a dimension independent way.

\subsubsection{Dimension independent setup}

Recall that, as we discussed before Eq. \eqref{LocalMode}, a generic field mode is specified by two dimensionless self-adjoint operators $\hat{V}$ and $\hat{W}$---both in the span of \mbox{$\{\hat\varphi(\mathsf{x})\ \vert \ \mathsf{x}\in\mathcal{M}\}$}---which fail to commute with each other as \mbox{$[\hat{V},\hat{W}]=\ii\hat{\openone}$}. Since we are interested here in finding the field mode whose reduced state is a) the least mixed possible, while at the same time, b) having support over only a compact region, $R\subset\mathcal{M}$, we will find these modes by a  minimization search of $\hat{V}$ and $\hat{W}$ satisfying \mbox{$\hat{V},\hat{W}\in\text{span}(\{\hat\varphi(\mathsf{x})\ \vert \ \mathsf{x}\in R\})$}. Thus, for some yet-to-be-specified measure of mixedness, $S$, we can define,
\begin{align}
S_\text{min}(R,\beta)\coloneqq  \text{inf}_{\hat{V},\hat{W}}\ \  &S(\hat{V},\hat{W},\hat{\rho})\\
\nonumber
\text{s.t.}\quad &[\hat{V},\hat{W}]=\ii\,\hat{\openone}\\
\nonumber
&\hat{V}\,\in\text{span}(\{\hat\varphi(\mathsf{x})\ \vert \  \mathsf{x}\in R\})\\
\nonumber
&\hat{W}\,\in\text{span}(\{\hat\varphi(\mathsf{x})\ \vert \  \mathsf{x}\in R\}).
\end{align}
We will call a pair of operators, $\hat{V}$ and $\hat{W}$, feasible if they satisfy the above three constraints.

For any unitarily invariant measure of mixedness (e.g., one minus purity, entropy, etc.), the pair of operators $\hat{V},\hat{W}$ which achieve this minimum\footnote{Technically speaking we are taking the infimum in the above expression. As such, there may not be any pair of operators $\hat{V},\hat{W}$ which achieve $S_\text{min}$. More carefully we should speak only of sequences of $(\hat{V},\hat{W})$ which in the limit achieve $S_\text{min}$.} will not be unique. Indeed, any Gaussian unitary\footnote{A Gaussian unitary is a unitary transformation generated by a quadratic Hamiltonian in the field and its conjugate. Alternatively, one can characterize Gaussian unitaries as those which map the field and its conjugate onto linear combinations of $\hat\varphi(\mathsf{x})$, $\hat\pi(\mathsf{x})$ and $\openone$. For further details, see our review of Gaussian Quantum Mechanics in Appendix \ref{GQM}.}  transformation (i.e., phase space displacement, rotation, squeezing or their combinations) which acts only within the $(\hat{V},\hat{W})$-mode will preserve both the feasibility of any pair $\hat{V}$, $\hat{W}$ and as well their mixedness. Let $\hat{X}$ and $\hat{Y}$ be a pair of dimensionless self-adjoint operators which are related to the generic $\hat{V}$ and $\hat{W}$ in this way. Using this Gaussian unitary freedom, we can impose further constraints on $\hat{X}$ and $\hat{Y}$, namely: $\langle\hat{X}\rangle_{\hat{\rho}}=0$, $\langle\hat{Y}\rangle_{\hat{\rho}}=0$, $\langle\{\hat{X},\hat{Y}\}\rangle_{\hat{\rho}}=0$ and $\langle\hat{X}^2\rangle_{\hat{\rho}}=\langle\hat{Y}^2\rangle_{\hat{\rho}}$. This amounts to symplectically diagonalizing the covariance matrix\footnote{The covariance matrix is a collection of the second moments of $\hat{V}$ and $\hat{W}$. For further details, see our review of Gaussian Quantum Mechanics in Appendix \ref{GQM}.} of the $(\hat{V},\hat{W})$-mode. For any $\hat{V}$ and $\hat{W}$, such $\hat{X}$ and $\hat{Y}$ exist.

Let us now take our measure of mixedness, $S(\hat{V},\hat{W},\hat{\rho})$, to be the symplectic eigenvalue\footnote{Roughly speaking, the symplectic eigenvalue corresponds to the ``occupation number'' of the mode $(\hat{V}, \hat{W})$ in state $\hat{\rho}$, and it provides a useful parameter to assess mixedness of general Gaussian states. For further details, see our review of Gaussian Quantum Mechanics in Appendix \ref{GQM}.}, $\nu\geq1$, which is associated with the covariance matrix of the $(\hat{V},\hat{W})$-mode. This mode's purity and entropy can be calculated straightforwardly from $\nu$. We note that $\nu=1$ corresponds to complete purity whereas $\nu\to\infty$ corresponds to complete mixedness. For $\hat{X}$ and $\hat{Y}$ (which recall obey the further constrains given above) this symplectic eigenvalue takes the form,
\begin{align}
\nu(\hat{X},\hat{Y},\hat{\rho})=\frac{1}{2}\big\langle\hat{X}^2+\hat{Y}^2\big\rangle_{\hat{\rho}}.
\end{align}
For the generic $\hat{V}$ and $\hat{W}$ (which may not obey these further constraints) the symplectic eigenvalue is the same as that of $\hat{X}$ and $\hat{Y}$, but the right-hand-side of the above equation may be larger. That is,
\begin{align}
\nu(\hat{X},\hat{Y},\hat{\rho})
=\nu(\hat{V},\hat{W},\hat{\rho})
\leq\frac{1}{2}\big\langle\hat{V}^2+\hat{W}^2\rangle_{\hat{\rho}}.
\end{align}
It is important to note that for any feasible $\hat{V}$ and $\hat{W}$ this inequality is saturated by some other feasible pair of operators (namely, $\hat{X}$ and $\hat{Y}$). Thus, identifying the minimally mixed mode with compact support over \mbox{$R\subset\mathcal{M}$} reduces to the following minimization problem:
\begin{align}
\nu_\text{min}(R,\beta)=\text{inf}_{\hat{V},\hat{W}}&\quad \frac{1}{2}\big\langle\hat{V}^2+\hat{W}^2\rangle_{\hat{\rho}}\\
\nonumber
\text{s.t.}&\quad[\hat{V},\hat{W}]=\ii\hat{\openone}\\
\nonumber
&\quad\hat{V},\hat{W}\in\text{span}(\{\hat\varphi(\mathsf{x})\ \vert \ x\in R\}).
\end{align}
We are thus in a sense looking for the minimum uncertainty mode with compact support over $R$. For notational convenience let us define this uncertainty measure as, 
\begin{align}
u(\hat{V},\hat{W},\hat\rho)\coloneqq\frac{1}{2}\big\langle\hat{V}^2+\hat{W}^2\rangle_{\hat{\rho}}.    
\end{align}
We will call this $u$ the objective function.

We claim that for any compact spacetime region, $R$, and any ambient field temperature, $\beta$, we have\footnote{Note that we are taking an infimum here such that $\nu_\text{min}=1$ does not imply that there exists a mode which itself is completely pure, i.e., $\nu=1$.} $\nu_\text{min}(R,\beta)=1$. That is, while (as was proved in \cite{ruep2021}) every local mode of a free quantum field in a thermal state is mixed ($\nu>1$), there are local modes in any compact region $R$ which are arbitrarily pure ($\nu\to1$). %We note again that while we take this claim to be true dimension-independently we will, for the sake of simplicity, only consider the 1+1 dimensional case in full detail. Proof of our claim in the 1+1 dimensional case will be given in the next subsection. For the rest of this subsection, let us proceed towards this proof in a dimension independent way.

We first note that by the isotony property, decreasing the size of $R$  never decreases $\nu_\text{min}$ because then strictly less $\hat{V}$ and $\hat{W}$ will be feasible. That is, 
\mbox{$\nu_\text{min}(R',\beta)\geq\nu_\text{min}(R,\beta)$} when \mbox{$R'\subset R$}. Since we know \mbox{$\nu_\text{min}\geq1$}, if we can prove that \mbox{$\nu_\text{min}(R',\beta)=1$}  for some \mbox{$R'\subset R$} then we will have proved that \mbox{$\nu_\text{min}(R,\beta)=1$}. This will be our proof strategy.

Assuming that the spacetime is flat, within any compact region, $R$, we can construct a subregion $R'$ as follows: Consider a point $\mathsf{x}_0$ in the interior of $R$. Take an inertial frame $(t,\bm x)$ with $\mathsf{x}_0$ as its origin. Choose the $t=0$ Cauchy surface  $\Sigma(t\!=\!0)$. Let $A_\ell$ be the set which is the intersection of $\Sigma(t\!=\!0)$ with an $\ell$-sized ball centered at $\mathsf{x}_0$. We take $R'=R_\ell\coloneqq D(A_\ell)$ to be the domain of dependence of $A_\ell$. Because $\mathsf{x}_0$ is in the interior of $R$, we will have $R_\ell\subset R$ for small enough $\ell$.

Next, recall that by using the time-slice property, we can identify any $\hat{V}$ and $\hat{W}$ within the algebra of $\hat\varphi$ on any thickened Cauchy surface. Equivalently, we can identify $\hat{V}$ and $\hat{W}$ within the algebra of $\hat\varphi$ and its conjugate momentum $\hat\pi$ on any (non-thickened) Cauchy surface. For the case of interest $\hat{V}$ and $\hat{W}$ are localized on some compact spacetime region, $R_\ell$. By construction when we move $\hat{V}$ and $\hat{W}$ onto the \mbox{$\Sigma(t\!=\!0)$} Cauchy surface, we will find that they are compactly supported over $A_\ell$. Thus we have,
\begin{align}
\nu_\text{min}(R_\ell,\beta)=\text{inf}_{\hat{V},\hat{W}}&\ \ \frac{1}{2}\big\langle\hat{V}^2+\hat{W}^2\rangle_{\hat{\rho}}\\
\nonumber
\text{s.t.}&\ \ [\hat{V},\hat{W}]=\ii\hat{\openone}\\
\nonumber
&\ \ \hat{V}\in\text{span}(\{\hat\varphi(\bm{x}),\hat\pi(\bm{x}) \ \vert \  \vert \bm{x}\vert<\ell\})\\
\nonumber
&\ \ \hat{W}\!\in\text{span}(\{\hat\varphi(\bm{x}),\hat\pi(\bm{x}) \ \vert \ \vert \bm{x}\vert<\ell\}).
\end{align}
where $\bm{x}$ is a vector in $\mathbb{R}^n$. Note that for notational convenience we have dropped the time dependence of the field operators: $\hat\varphi(\bm{x})=\hat\varphi(0,\bm{x})$ and $\hat\pi(\bm{x})=\hat\pi(0,\bm{x})$. We will now prove that $\nu_\text{min}(R_\ell,\beta)=1$ for every lengthscale $\ell$ and every ambient field temperature $\beta$. As we have discussed above, since we can find some \mbox{$R_\ell$} within any region $R$ this is sufficient to prove $\nu_\text{min}(R,\beta)=1$ for every $R$ and $\beta$.

Our proof method will be as follows: We will provide a sequence of feasible operators, $\hat{V}$ and $\hat{W}$, which achieve the limit \mbox{$u=\frac{1}{2}\big\langle\hat{V}^2+\hat{W}^2\rangle_{\hat{\rho}}\to1$}, where we dropped the arguments of the objective function $u(\hat{V},\hat{W},\hat\rho)$ for notational convenience. In particular, our sequence will have $\hat{V}$ supported only by $\hat\varphi(\bm{x})$ and $\hat{W}$ supported only by $\hat\pi(\bm{x})$ as,
\begin{align}\label{VandW}
\hat{V}&=\int \dd^n \bm{x}\,
v(\bm{x})\,\hat\varphi(\bm{x}), & 
\hat{W}&=\int \dd^n \bm{x}\,
w(\bm{x})\,\hat\pi(\bm{x}),
\end{align}
for some real functions $v(\bm{x})$ and $w(\bm{x})$. In these terms, the compact support constraint is just that both $v(\bm{x})$ and $w(\bm{x})$ have support only over $\bm{x}$ with $\vert\bm{x}\vert<\ell$. The commutation constraint for $\hat{V}$ and $\hat{W}$ can be written in terms of $v(\bm{x})$ and $w(\bm{x})$ as, 
\begin{align}\label{CommConstraint}
[\hat{V},\hat{W}]=\ii\hat{\openone}
\Leftrightarrow
\int\dd^n \bm{x}\, v(\bm{x}) \,w(\bm{x})= 1.
\end{align}
Finally, the objective function in terms of $v(\bm{x})$ and $w(\bm{x})$ is,
\begin{align}
u=&\frac{1}{2}\int\!\!\!\int\dd^n\bm{x}\,\dd^n\bm{y}\ \langle\hat\varphi(\bm{x})\hat\varphi(\bm{y})\rangle_{\hat\rho}\ v(\bm{x}) v(\bm{y})\,\\
\nonumber
+&\frac{1}{2}\int\!\!\!\int\dd^n \bm{x}\,\dd^n \bm{y}\ \langle\hat\pi(\bm{x})\hat\pi(\bm{y})\rangle_{\hat\rho}\ w(\bm{x}) w(\bm{y}),
\end{align}
where the thermal two-point functions are well-known for free scalar fields~\cite{Weldon}. Written in terms of the Fourier transforms of $v(\bm{x})$ and $w(\bm{x})$, the objective function then takes the form
\begin{align}\label{UdimInd}
    u = \dfrac{1}{4}\Big[&\int\dd^n k\dfrac{1}{\omega_{\bm{k}}}\coth\left(\beta\omega_{\bm{k}}/2\right)\abs{\Tilde{v}(\bm{k})}^2\nonumber\\
    +&\int\dd^n k\,\,\omega_{\bm{k}}\coth\left(\beta\omega_{\bm{k}}/2\right)\abs{\Tilde{w}(\bm{k})}^2\Big].
\end{align}
For $n\geq 2$ spatial dimensions, we will take our sequence to have  $v(\bm{x})$ and $w(\bm{x})$ spherically symmetric. In this case the objective function~\eqref{UdimInd} simplifies greatly, since the Fourier transforms $\Tilde{v}(\bm{k})$ and $\Tilde{w}(\bm{k})$ will only depend on $\abs{\bm{k}}$, and the angular part of the integral in~\eqref{UdimInd} can be readily computed. In this case, Eq.~\eqref{UdimInd} then becomes
\begin{align}\label{UdimInd2}
u = \dfrac{\pi^{n/2}}{2\Gamma\left(\frac{n}{2}\right)}\Big[&\int_{0}^{\infty}\!\!\!\dd k \dfrac{k^{n-1}}{\omega_{\bm{k}}}\coth\left(\dfrac{\beta\omega_k}{2}\right)\abs{\Tilde{v}(k)}^2 \nonumber \\
+&\int_{0}^{\infty}\!\!\!\dd k\, k^{n-1}\omega_{\bm{k}}\coth\left(\dfrac{\beta\omega_k}{2}\right)\abs{\Tilde{w}(k)}^2\Big],
\end{align}
where $\omega_{\bm{k}}=\vert \bm{k}\vert$ is the field's dispersion relation and \mbox{$k\equiv|\bm k|$} is just shorthand for the modulus of the wave vector. It should be noted that the factor of $k^{n-1}$ above makes both integrals convergent in the IR for for $n\geq 3$. For $n=1$ or $n=2$ we will need to choose $\Tilde{v}(k)$ and $\Tilde{w}(k)$ to go to zero sufficiently quickly as $k\to0$ to regularize the IR. For all $n$ we need $\Tilde{v}(k)$ and $\Tilde{w}(k)$ to go to zero sufficiently quickly as $k\to\infty$ to regularize the UV.

In the following subsection we consider the flat 1+1 dimensional case and will produce a sequence of $v(\bm{x})$ and $w(\bm{x})$ with $u\to1$ in the limit. However, at the end of the next subsection we will discuss the prospects of generalizing our proof to higher dimensions and to curved spacetimes.

%While we take this claim to be true dimension-independently, for simplicity we will consider only the 1+1 dimensional case in full technical detail. Our proof in the 1+1 dimensional case will be given in the next subsection. Following this we will comment on how this proof might be adapted to higher dimensional cases. We believe that our the results can likely be generalized to higher dimensional settings quite easily. As such, in this section we will state the problem in a dimension independent way.

\subsubsection{1+1 dimensional proof}
For simplicity we will now restrict our attention to the 1+1 dimensional case. In this case the objective function \eqref{UdimInd} simplifies to,
\begin{align}
u
&=\frac{1}{2}\int_{-\infty}^\infty\dd k\, \frac{1}{\omega_k}\,\text{coth}(\beta\omega_k/2)\, \vert \tilde{v}(k)\vert^2\\
\nonumber
&+\frac{1}{2}\int_{-\infty}^\infty\dd k\, \omega_k\,\text{coth}(\beta\omega_k/2)\, \vert \tilde{w}(k)\vert^2,
\end{align}
In order to show that $\nu_\text{min}(R_\ell,\beta)=1$ we will now present a sequence of functions $v_{m,\kappa}(x)$ and $w_{m,\kappa}(x)$ which in the limit have $u\to1$ while satisfying the above discussed constraints. Namely, each pair of $v_{m,\kappa}(x)$ and $w_{m,\kappa}(x)$ will obey the commutation condition \eqref{CommConstraint} with $n=1$ and will be supported only over $x\in[-\ell,\ell]$.

Consider the two-parameter family of functions,
\begin{align}\label{vMKwMKdef}
v_{m,\kappa}(x)&=\sqrt{\omega_\kappa}\ z_{m,\kappa}(x),\\
w_{m,\kappa}(x)&=\frac{1}{\sqrt{\omega_\kappa}}\,z_{m,\kappa}(x),
\end{align}
where
\begin{align}\label{zMKdef}
z_{m,\kappa}(x) = A_{m,\kappa}\,\partial_x^2\left[\text{cos}(\kappa x)\,B_m(x/\ell)\right]
\end{align}
is parametrized by a real number $\kappa >0$ and an integer $m\geq3$. The prefactor $A_{m,\kappa}$ ensures that $z_{m,\kappa}(x)$ is $L^2$ normalized. $B_m(s)$ are the B-spline functions~\cite{WolframBSpline} which result from repeated self-convolution of the rectangle function,
\begin{align}\label{BMdef}
B_m(s) \coloneqq \Pi^{\ast m}(m\,s)
\end{align}
where  $\Pi(s)$ is the rectangle function over the domain \mbox{$s\in[-1,1]$}, the operator $\ast$ indicates convolution, and the superscript $\ast m$ indicates repeated self-convolution. For instance, $B_1(s)$ is the rectangle function over \mbox{$s\in[-1,1]$}, and $B_2(s)$ is the triangle function over \mbox{$s\in[-1,1]$}. In general, $B_m(s)$ is supported over \mbox{$s\in[-1,1]$} and composed piecewise of $m$ polynomials of degree $m-1$.

These are convenient functions to investigate because a) they ensure that $\hat{V}$ and $\hat{W}$ are automatically feasible, and b) their Fourier transforms take a convenient form. To see the feasibility of $\hat{V}$ and $\hat{W}$, first note that since $z_{m,\kappa}(x)$ is $L^2$ normalized, $v_{m,\kappa}(x)$ and $w_{m,\kappa}(x)$ satisfy the commutation constraint, Eq. \eqref{CommConstraint}. Secondly note that $B_m(x/\ell)$ is compactly supported over \mbox{$x\in[-\ell,\ell]$}. Moreover, multiplying by $\text{cos}(\kappa x)$ and taking two derivatives does not change the function's support. As such $z_{m,\kappa}(x)$---and therefore also $v_{m,\kappa}(x)$ and $w_{m,\kappa}(x)$---are compactly supported over \mbox{$x\in[-\ell,\ell]$}. 

\begin{figure}
\includegraphics[width=0.5\textwidth]{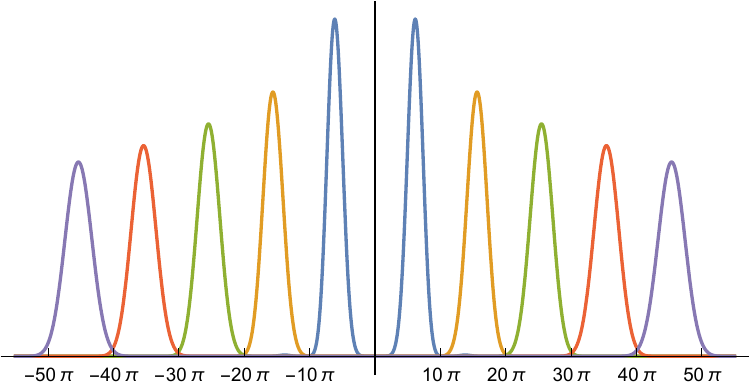}
\caption{The power spectrum, $\vert \tilde{z}_{m,\kappa}(k)\vert^2$, given by \eqref{PowerSpectrum} is shown for several values of $m$ and $\kappa$. The peaks are paired mirror-wise. From inside to outside we have ($m=3$, $\kappa=5\pi$), ($m=4$, $\kappa=15\pi$), ($m=5$, $\kappa=25\pi$), ($m=6$, $\kappa=35\pi$), and ($m=7$, $\kappa=45\pi$) with $\ell=1$ throughout. As $\kappa$ increases the peaks are displaced farther from the origin. As $m$ is increased the polynomial tails of the distributions are suppressed. Away from the peaks $\vert\tilde{z}_{m,\kappa}(k)\vert^2$ decays polynomially as $\vert k\vert^{4-2m}$. Increasing $m$ also widens the peaks. The peak width is proportional to $m$.}\label{Figurezk2}
\end{figure}

The Fourier transform of $z_{m,\kappa}(x)$ is given by
\begin{align}
\nonumber 
\tilde{z}_{m,\kappa}(k)
&=k^2\,C_{m,\kappa}\! \big[\!\left(\delta(k-\kappa)+\delta(k+\kappa)\right)\ast S_m(k\ell)\big],\\
\label{PowerSpectrum}
S_m(r)&\coloneqq\left[\frac{\text{sin}(r/m)}{r/m}\right]^m.
\end{align}
where the $L^2$ normalizing prefactor, $A_{m,\kappa}\to C_{m,\kappa}$, has absorbed some stray constants. $S_m(k\ell)$ is proportional to the Fourier transform of $B_m(x/\ell)$. The convolution, $\ast$, with the two delta functions accounts for the multiplication by $\text{cos}(\kappa x)$. Finally, the $k^2$ appearing above expression is a consequence of the two derivatives taken in Eq. \eqref{zMKdef}. To anticipate what is to come, this factor of $k^2$ helps us regulate the IR divergences present in \eqref{UdimInd} for $n=1$. Fig.~\ref{Figurezk2} shows $\vert \tilde{z}_{m,\kappa}(k)\vert^2$ for values of increasing $m$ and $\kappa$.

As Fig.~\ref{Figurezk2} shows, $\vert\tilde{z}_{m,\kappa}(k)\vert^2$ has two peaks located roughly at $k=\pm \kappa$. By increasing $\kappa$ we can push these peaks as far apart as we like. Away from these peaks $\vert\tilde{z}_{m,\kappa}(k)\vert^2$ decays polynomially as $\vert k\vert^{4-2m}$. By increasing $m$ we can suppress these tails as much as we like. However, increasing $m$ does have the effect of widening both peaks: note $S_m(k\ell)$ has its first zeroes at $k=\pm m\pi/\ell$. As we will now discuss, by increasing $\kappa$ and $m$ we can make the objective function, $u=\frac{1}{2}\langle\hat{V}^2+\hat{W}^2\rangle_{\hat\rho}$, as near to one as we wish.

Written in terms of the Fourier transform of $z_{m,\kappa}(x)$, the objective function is
\begin{align}
\nonumber
u&=\frac{1}{2}\int_{-\infty}^\infty\dd k \left(\frac{\omega_\kappa}{\omega_k}+\frac{\omega_k}{\omega_\kappa}\right)\,\text{coth}(\beta\omega_k/2)\, \vert \tilde{z}_{m,\kappa}(k)\vert^2\\
&\eqqcolon\left\langle\frac{1}{2} \left(\frac{\omega_\kappa}{\omega_k}+\frac{\omega_k}{\omega_\kappa}\right)\,\text{coth}(\beta\omega_k/2)\right\rangle_{\vert \tilde{z}_{m,\kappa}(k)\vert^2}.
\end{align}
First it should be noted that this integral is finite. The IR divergences coming from \mbox{$1/\omega_k\to\infty$} and \mbox{$\text{coth}(\beta\omega_k/2)\to\infty$} as \mbox{$k\to0$} are controlled by $k^2$ appearing in $\tilde{z}_{m,\kappa}(k)$. Moreover, the UV divergence coming from \mbox{$\omega_k\to\infty$} as \mbox{$k\to\infty$} is controlled by the polynomial tails of $\tilde{z}_{m,\kappa}(k)$ so long as $m\geq3$.

Let us discuss the asymptotic behaviour of the sequence of functions $v_{m,\kappa}(x)$ and $w_{m,\kappa}(x)$: As we will now show, by increasing both $\kappa$ and $m$ we can bring the objective function as near to $1$ as we like.  First we note that by taking $\kappa$ large enough we can move an arbitrary amount of the support of $\vert \tilde{z}_{m,\kappa}(k)\vert^2$ into a regime where $\text{coth}(\beta\omega_k/2)\approx1$. Moreover, this approximation can be made arbitrarily accurate by increasing $\kappa$. Making this approximation the objective function becomes,
\begin{align}
u\approx\frac{1}{2}\left\langle\frac{\omega_\kappa}{\omega_k}+\frac{\omega_k}{\omega_\kappa}\right\rangle_{\vert \tilde{z}_{m,\kappa}(k)\vert^2}.
\end{align}
Note that for large enough $\kappa$ we can also approximate both $\omega_k/\omega_\kappa$ and $\omega_\kappa/\omega_k$ as linear functions of $k$ in the neighborhoods around $k=\pm \kappa$. For any fixed width around our two peaks we can also make these linear approximations arbitrarily accurate within those regions by increasing $\kappa$. Thus, even if $m$ is large such that the peaks around $k=\pm \kappa$ are wide, we can always move to larger $\kappa$ to justify this approximation over the peak's width. 

Finally, note that the average of any linear function, $L(x)$, with respect to any weight function\footnote{I.e., a   probability distribution.}, $p$, is just the value of the function at the center of that profile. That is, \mbox{$\langle L(x)\rangle_p = L\big(\langle x\rangle_p\big)$}. Thus recalling that we can approximate both $\omega_k$ and $1/\omega_k$ linearly, we have that for large enough $\kappa$,
\begin{align}
\left\langle\omega_k\right\rangle_{\vert \tilde{z}_{m,\kappa}(k)\vert^2}&\approx\omega_\kappa\\
\left\langle\frac{1}{\omega_k}\right\rangle_{\vert \tilde{z}_{m,\kappa}(k)\vert^2}&\approx\frac{1}{\omega_\kappa}
\end{align}
to arbitrary accuracy. Therefore in the limit $\kappa\to\infty$ we have \mbox{$u\approx1$} to arbitrary accuracy. This establishes that \mbox{$u\to1$} as claimed above. 

We have thus shown for a free 1+1 dimensional thermal quantum field in a flat spacetime that within any compact spacetime region $R\subset\mathcal{M}$ we can find a local mode which is arbitrarily pure. To extend this proof to higher dimensions, one simply needs to design $v(x)$ and $w(x)$ with $u\to1$ for each $n$. Concerning curved spacetimes, since this proof works primarily in the UV of the theory (i.e., large $k$ and small $\ell$) it should also be true in curved spacetimes at least at scales where the spacetime looks flat enough and as long as the spacetime has enough symmetry for the definitions of a vacuum and thermal field states to be unambiguous.

\subsection{Realistic Levels of Mixedness}
In the previous subsection we have seen that local modes can be arbitrarily pure regardless of the region size $\ell$ and the ambient field temperature $\beta$. However, constructing these arbitrarily pure modes required us to go into the deep UV of our theory. What level of mixedness can we expect for more realistic local modes? In this subsection we will provide a quantitative analysis of the mixedness of a few local mode profiles in $1+1$ and $3+1$ dimensions. As we will see the dependence of the local mode's purity on its shape, size, and the ambient temperature all indicate that local field modes of a free theory are not good models for experimentally realistic localized probes.

We will consider local modes specified by $\hat{V}$ and $\hat{W}$ given by \eqref{VandW} for some $v(\bm{x})$ and $w(\bm{x})$. In particular, we take $v(\bm{x})$ and $w(\bm{x})$ to be spherically symmetric functions which are supported only over $\bm{x}$ with $\vert\bm{x}\vert<\ell$ and satisfy the commutation constraint, Eq.~\eqref{CommConstraint}. For a thermal state we know that $\langle\{\hat{V},\hat{W}\}\rangle_{\hat\rho}=0$ as well as $\langle\hat{V}\rangle_{\hat\rho}=0$ and $\langle\hat{W}\rangle_{\hat\rho}=0$. For such $\hat{V}$ and $\hat{W}$, the symplectic eigenvalue associated with this mode is given by,
\begin{align}\label{NuSqrtDef}
\nu=\sqrt{\langle\hat{V}^2\rangle_{\hat\rho}\,\langle\hat{W}^2\rangle_{\hat\rho}}    
\end{align}
where for spatial dimension $n=1$ we have, formally,
\begin{align}
\langle\hat{V}^2\rangle_{\hat\rho}
&=\frac{1}{2}\int_{-\infty}^\infty\dd k\, \frac{1}{\omega_k}\,\text{coth}(\beta\omega_k/2)\, \vert \tilde{v}(k)\vert^2\\
\nonumber
\langle\hat{W}^2\rangle_{\hat\rho}&=\frac{1}{2}\int_{-\infty}^\infty\dd k\, \omega_k\,\text{coth}(\beta\omega_k/2)\, \vert \tilde{w}(k)\vert^2,
\end{align}
and for spatial dimension $n\geq 2$ we have,
\begin{align}
\langle\hat{V}^2\rangle_{\hat\rho} &= \dfrac{\pi^{n/2}}{2\Gamma\left(\frac{n}{2}\right)}\int_{0}^{\infty}\!\!\!\dd k \dfrac{k^{n-1}}{\omega_{\bm{k}}}\coth\left(\dfrac{\beta\omega_{\bm{k}}}{2}\right)\abs{\Tilde{v}(k)}^2\\
\nonumber
\langle\hat{W}^2\rangle_{\hat\rho}&=\dfrac{\pi^{n/2}}{2\Gamma\left(\frac{n}{2}\right)}\int_{0}^{\infty}\!\!\!\dd k\, k^{n-1}\omega_{\bm{k}}\coth\left(\dfrac{\beta\omega_{\bm{k}}}{2}\right)\abs{\Tilde{w}(k)}^2,
\end{align}
where $\omega_{\bm{k}}$ is the field's dispersion relation and $k\equiv\vert\bm{k}\vert$ is shorthand for the modulus of the wavevector, $\bm{k}$. 

As we noted following equation \eqref{UdimInd2}, for $n=1$ and $n=2$ there may be an IR divergence in $\langle\hat{V}^2\rangle_{\hat\rho}$ if $\abs{\Tilde{v}(k)}^2$ does not go to zero quickly enough as $k\to0$. When needed, we will consider two kinds of IR regularization: 1) we can take the field to be massive with a  mass $M$. In this case the dispersion relation becomes \mbox{$\omega_{\bm{k}}=\sqrt{\vert\bm{k}\vert^2+M^2}$}. Alternatively, 2) we can take the field to be in a Dirichlet cavity of size $L$. For $n=1$ spatial dimensions this discretizes the allowed wavenumbers as $k_j=j\pi/L$ for $j=1,2,3,\dots$. In either case, our primary interest will be in the $M\to0$ and $L\to\infty$ limits of the regulators, when they exist.

\subsubsection{Mixedness of select 1+1 and 3+1 dimensional profiles}\label{SectionProfiles}
\begin{figure}
\includegraphics[width=0.5\textwidth]{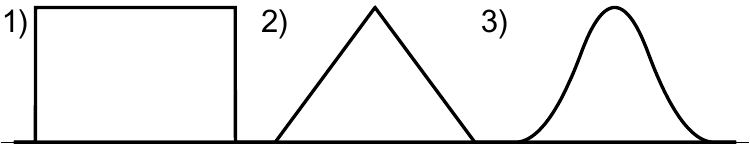}
\includegraphics[width=0.5\textwidth]{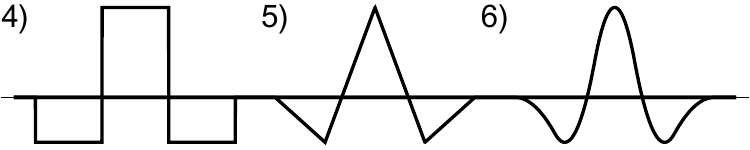}
\includegraphics[width=0.5\textwidth]{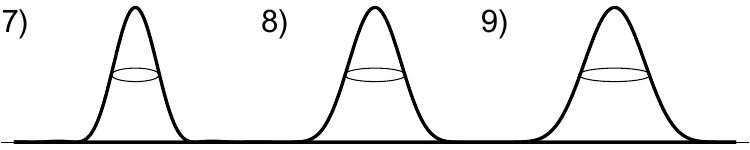}
\caption{The profiles of the local modes considered in this paper. Specifically, profiles 1, 2 and 3 show  Eq.~\eqref{BModes} with $m=1, 2, 3$. Likewise profiles 4, 5, and 6 show Eq.~\eqref{D2Bmodes} with $m=4,5,6$. Unlike the previous profiles, the final three profiles are 3 dimensional. Specifically, profiles 7, 8 and 9 show Eq.~\eqref{B3D} with $m=2, 3, 4$. }\label{FigureZ}
\end{figure}

As a first example, let us consider the local modes in a 1+1 dimensional flat spacetime. Specifically, let us consider profiles 1 through 6 shown in Figure \ref{FigureZ}. Profiles 1, 2 and 3 are given by
\begin{align}\label{BModes}
v(x)=w(x)=A_m \, B_m(x/\ell)
\end{align}
with $m=1,2,$ and $3$ where $B_m(s)$ are the B-spline functions defined in Eq.~\eqref{BMdef}. The constants $A_m\in\mathbb{R}$ are chosen so that the functions are $L^2$-norm one. As we will see, the mixedness of these modes with these profiles presents an IR divergence as we remove the IR regulator (i.e., taking either $m\to0$ or $L\to\infty$). Profiles 4, 5 and 6 in Figure \ref{FigureZ} are given by
\begin{align}\label{D2Bmodes}
v(x)=w(x)=C_m \, \partial_x^2\,B_{m-1}(x/\ell)
\end{align}
with $m=4,5,$ and $6$ where the $C_m\in\mathbb{R}$ are chosen such that the functions have $L^2$-norm of one. As we will see, the modes with these `derivative' profiles do not have an IR divergence as we take either $M\to0$ or $L\to\infty$.

\begin{figure*}
\includegraphics[width=0.32\textwidth]{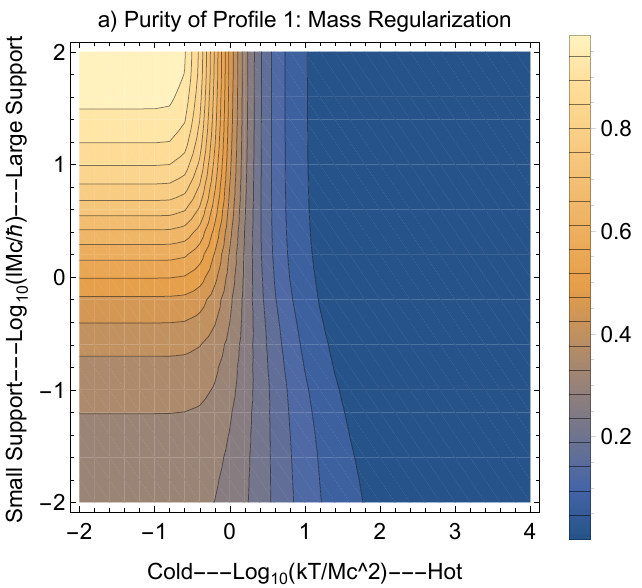}
\includegraphics[width=0.32\textwidth]{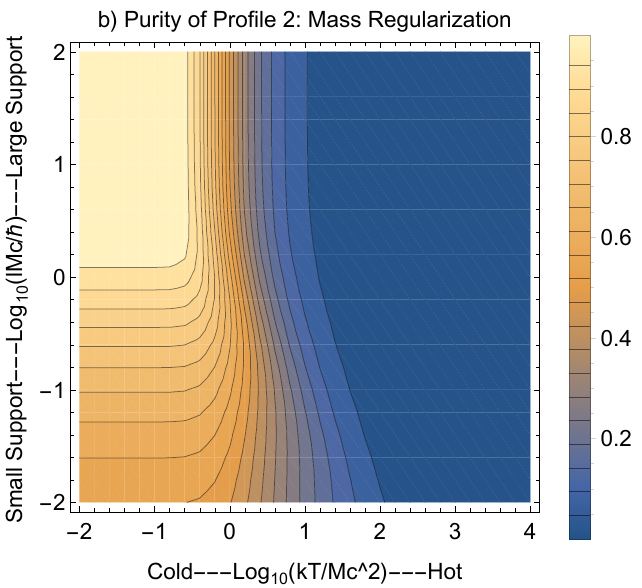}
\includegraphics[width=0.32\textwidth]{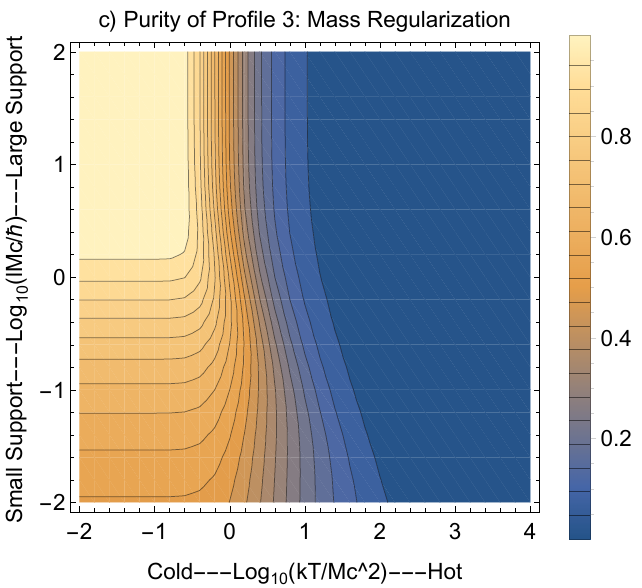}\\
\includegraphics[width=0.32\textwidth]{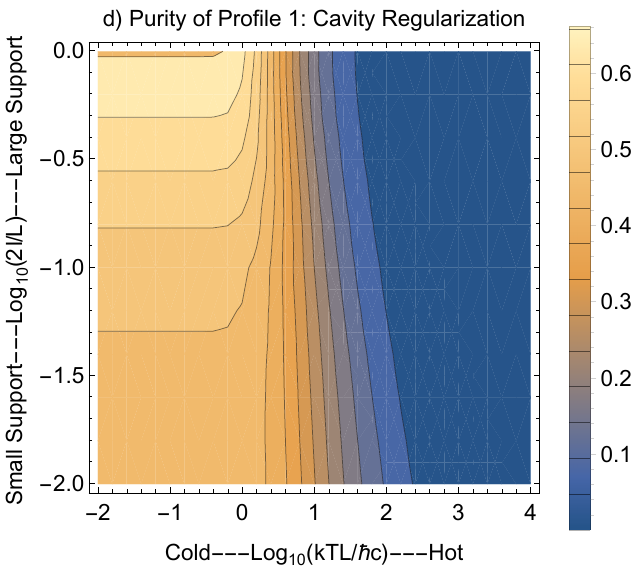}
\includegraphics[width=0.32\textwidth]{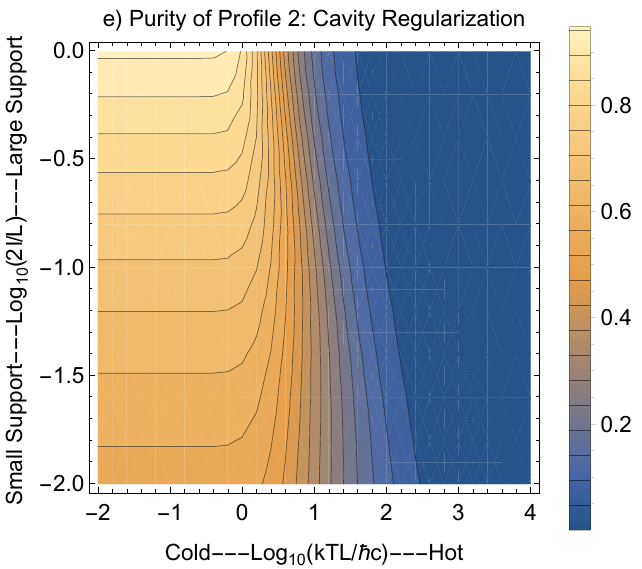}
\includegraphics[width=0.32\textwidth]{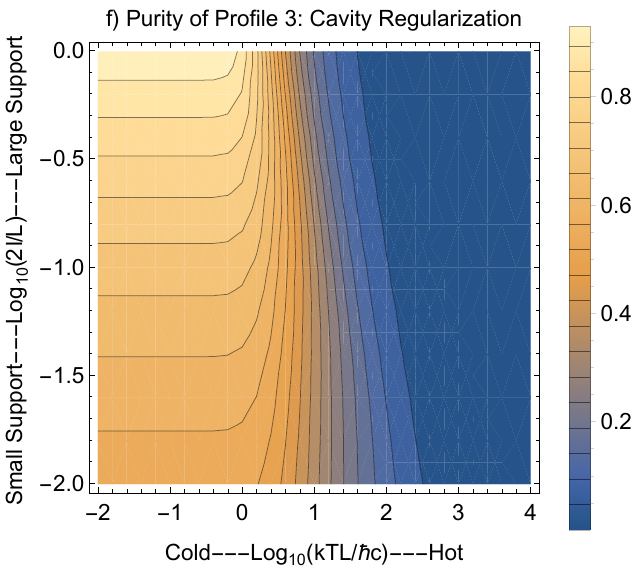}
\caption{The purity of select local modes as we vary the size of their support (vertical axis) and the ambient temperature of the field (horizontal axis). The top row of subfigures has an IR regulator introduced by means of a field mass $M$ to which the support size $\ell$ and ambient temperature $T$ are compared. The bottom row of subfigures has an IR regulator introduced by means of cavity walls a distance $L$ apart fom each other. The columns show the purity of the local mode given by profiles 1, 2 and 3 in Figure \ref{FigureZ}.}\label{PurityFiguresSeries1}
\end{figure*}

Let us first consider profiles 1, 2 and 3. The purity \mbox{(i.e, $P\coloneqq\text{Tr}(\hat\rho_\text{local}^2)=1/\nu$)} of these modes is shown in Fig.~\ref{PurityFiguresSeries1}. On the top row the IR regularization is set by the field mass $M$ and on the bottom row it is set by a cavity length scale $L$. Let us consider the massive case first. 

As the top row of Fig.~\ref{PurityFiguresSeries1} shows, for these modes we have high purity only when both the temperature is much less than the mass scale (when $k_\text{B}T\ll Mc^2$, on the left side of each figure) and when the mode's support is much larger than the mass scale (when $\ell\gg \hbar/M c$, on the top of each figure). In the massless limit $M\to0$ we move downward and rightward on these figures. Due to the IR divergence in $\langle\hat{V}^2\rangle_{\hat\rho}$ we have $P\to0$ in the $M\to0$ limit for these modes.

Note that for temperatures at or above the mass scale, $kT\gtrsim Mc^2$, the contour lines become nearly vertical, indicating that these modes' purities are nearly independent of the support size in this regime. Conversely, for temperatures below the mass scale, $kT\lesssim Mc^2$, the contour lines are nearly horizontal such that the modes' purities are nearly independent of the ambient temperature in this regime. Thus (at least for these modes, and with this IR regularization) we have an idea of when the modes mixedness is dominated by temperature effects and when it is dominated by localization effects. 

Where can we expect a typical laboratory situation to lie on these figures? As a quasi-realistic\footnote{As we will discuss more  in depth later, pretending to accurately model any kind of atomic probe by using a (temporarily) localized  free field theory mode is problematic and provides misleading conclusions about purity.} example, consider a local mode of atomic size, $\ell=53\text{ pm}$ (Bohr radius), in a field with the mass of an electron, $M= 511\text{ keV/$c^2$}$ (electron mass) and a temperature of $T=4.2\,\text{K}$ (liquid Helium temperature). In this scenario we have $\ell M c/\hbar = 137$ and $k_\text{B}T/Mc^2 = 7\times10^{-10}$. This places us in the top left corner of these figures, where we have very high purity and near independence from temperature. Indeed, one might expect that in practice we will always be on the far-left sides of these figures where localization dominates; one rarely deals with temperatures larger than the mass scale in practice. However it is worth recalling that as we remove the IR regularization $M\to0$, we move both downward and rightward in these figures. That is, we move into the regime where mixedness due to temperature is dominant (on the right side) and into lower purities (downwards). We will hold the interpretation of these results for the moment: before extracting conclusions from this study, we will first see that things actually change substantially when we consider other IR regularizations and other mode profiles.

\begin{figure*}
\includegraphics[width=0.32\textwidth]{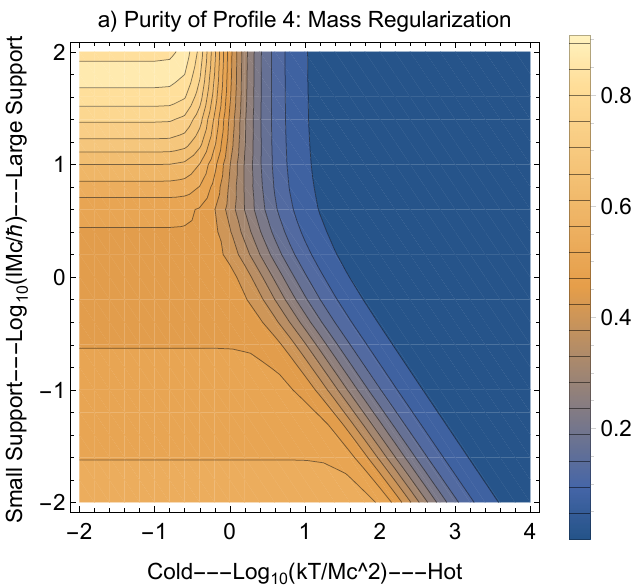}
\includegraphics[width=0.32\textwidth]{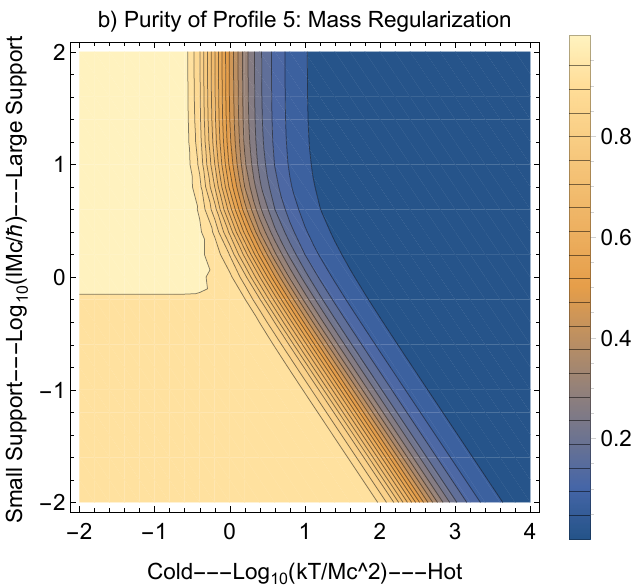}
\includegraphics[width=0.32\textwidth]{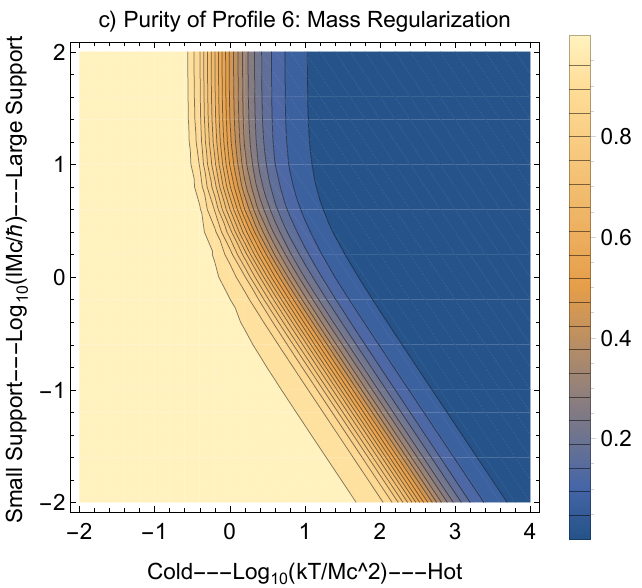}
\includegraphics[width=0.32\textwidth]{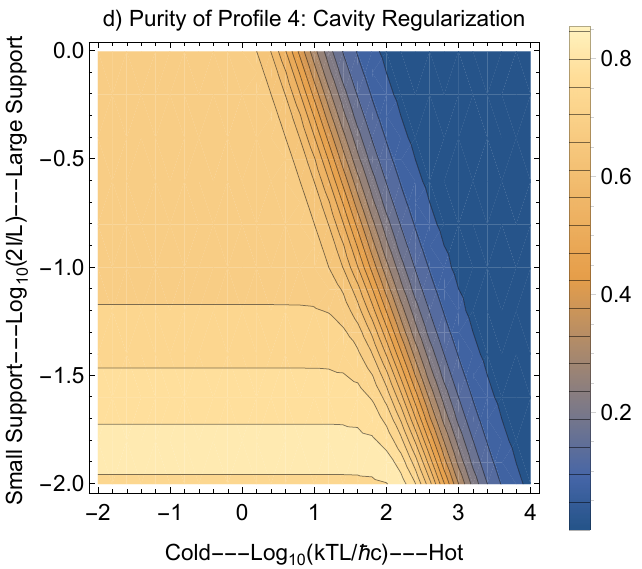}
\includegraphics[width=0.32\textwidth]{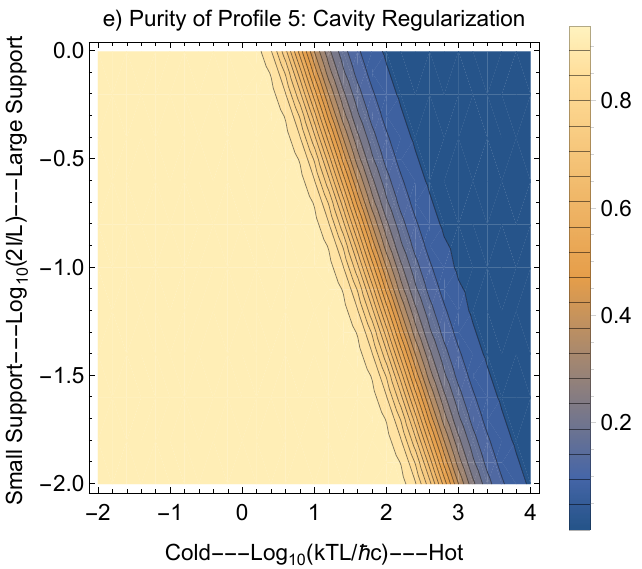}
\includegraphics[width=0.32\textwidth]{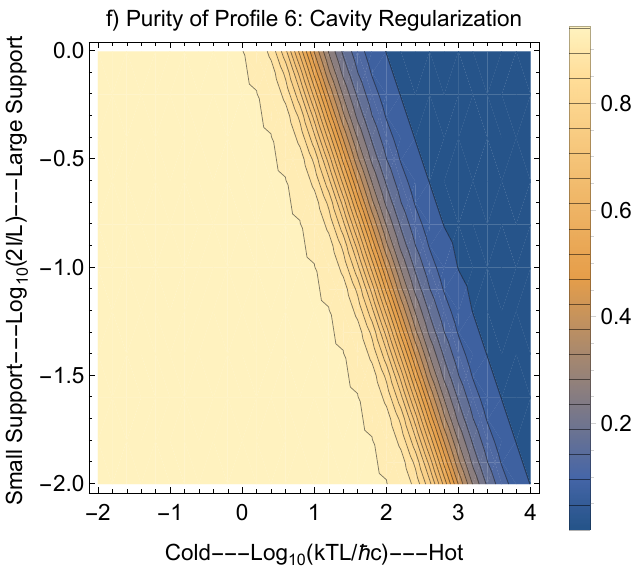}
\caption{The purity of select local modes as we vary the size of their support (vertical axis) and the ambient temperature of the field (horizontal axis). The top row of subfigures has an IR regulator introduced by means of a field mass $M$ to which the support size $\ell$ and ambient temperature $T$ are compared. The bottom row of subfigures has an IR regulator introduced by means of cavity walls a distance $L$ apart fom each other. The columns show the purity of the local mode given by profiles 4, 5 and 6 in Figure \ref{FigureZ}.}\label{PurityFiguresSeries2}
\end{figure*}

Let us consider the same three mode profiles, but now massless, $M=0$, and an IR regularization introduced by placing the field in a cavity of size $L$. The purity of the first three local modes with this IR regularization are shown on the bottom row of Fig.~\ref{PurityFiguresSeries1}. As these figure shows, for these modes we have high purity only when both the temperature is much less than the cavity scale (when $k_\text{B}T\ll \hbar c/L$, on the left side of the figure) and when the mode's support is comparable to the cavity size (when $\ell\sim L$ at the top of the figure). In the limit $L\to0$ we move downward and rightward on this figure. As before due to the IR divergence in $\langle\hat{V}^2\rangle_{\hat\rho}$ we have in the $L\to0$ limit $\mu\to0$ for these modes.

We can again identify in which regimes the mixedness is dominated by localization and when it is dominated by temperature. For this IR regularization the mixedness is localization-dominated (i.e., the purity is nearly independent of temperature) when the ambient temperature is less than the scale set by the cavity size. When the temperature is larger than this scale the mixedness is temperature-dominated (i.e., the purity is nearly independent of the support size). 

We can again ask where can we expect a typical situation to lie in these figures. As a quasi-realistic\footnote{Again, we warn strongly against using a local mode of a free field as a realistic model of an atomic probe. We will discuss this in more detail later.} example, consider a local mode of atomic size, $\ell=53\text{ pm}$, in a room-sized cavity, $L= 1\text{ m}$ and a temperature of $T=4.2\,\text{K}$. In this scenario we have $\ell/L = 5\times10^{-11}$ and $k_\text{B}T\,L/\hbar c = 1834$. This places us in the bottom right corner of these figures, where we have very low purity and near independence from the support size $\ell$. Indeed, one might expect that in practice we will always be on the far-right temperature dominated sides of these figures; as $L\to\infty$ every temperature will eventually have $k_\text{B}T\,L/\hbar c\gg1$. 

It should be stressed that the qualitative features of the above discussion (regularization via $M$ or $L$) are limited to the particular modes we are studying here---namely, Eq.~\eqref{BModes} with $m=1,2$ and $3$. As will now discuss, the situation changes qualitatively when we consider the modes given by profiles 4, 5 and 6 in Figure \ref{FigureZ} which do not need IR regularization.

\begin{figure*}
\includegraphics[width=0.32\textwidth]{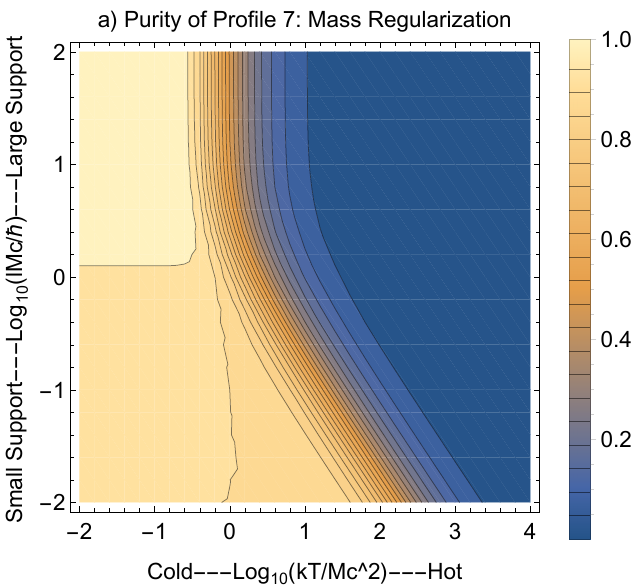}
\includegraphics[width=0.32\textwidth]{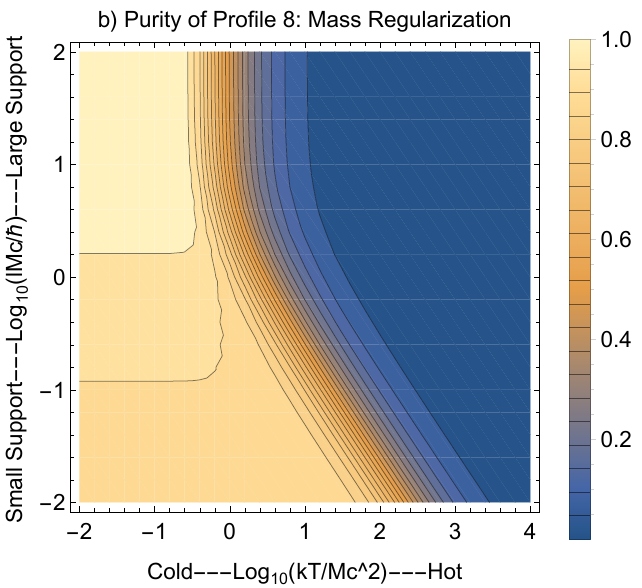}
\includegraphics[width=0.32\textwidth]{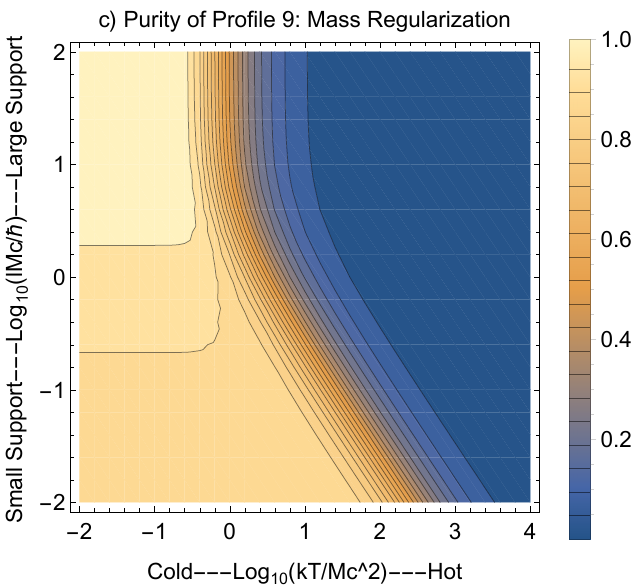}
\caption{The purity of select local modes as we vary the size of their support (vertical axis) and the ambient temperature of the field (horizontal axis). Each subfigure has an IR regulator introduced by means of a field mass $M$ to which the support size $\ell$ and ambient temperature $T$ are compared. The subfigures show the purity of the local mode given by profiles 7, 8 and 9 in Figure \ref{FigureZ}.}\label{PurityFiguresSeries3}
\end{figure*}

\begin{figure*}
\includegraphics[width=0.45\textwidth]{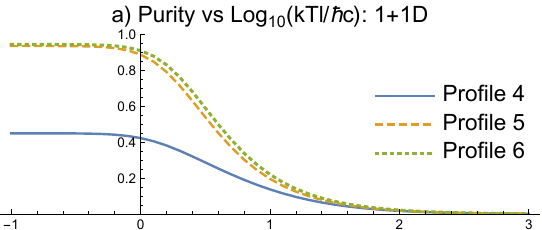}\qquad
\includegraphics[width=0.45\textwidth]{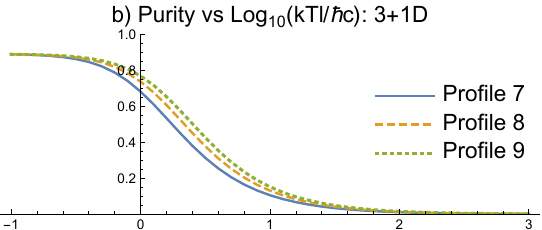}
\caption{The purity of three different local modes is shown without an IR regulator. In subfigure a) the field is 1+1 dimensional and in b) the field is 3+1 dimensional. In figure a) The three lines show the purity of the local mode given by profiles 4, 5 and 6 in Figure \ref{FigureZ}. b) The three lines show the purity of the local mode given by profiles 7, 8 and 9 in Figure \ref{FigureZ}.}\label{LimitPurity}
\end{figure*}

The purity of these modes is shown in Fig.~\ref{PurityFiguresSeries2}. On the top row we consider a field of mass $M$ for comparison (although no IR regularization is needed in this case). On the bottom row this IR regularization is set by the cavity length $L$. As before,  $M\to0$ or $L\to\infty$ moves us downwards and rightwards in each figure. The key qualitative difference between the purity of these three modes and those previously considered is that now the purity is well-defined as we take the massless or infinite length limits. The behavior of these modes purities when $M=0$ and $L\to \infty$ is shown in Fig.~\ref{LimitPurity}a.

The intuition from both Fig.~\ref{PurityFiguresSeries2} and Fig.~\ref{LimitPurity}a is that the purity of these modes increases as we lower the ambient field temperature, moving leftward within each figure. However, perhaps counter-intuitively, we have increasing purity as we make the support of the mode smaller in some regimes (moving downward in each figure). To see why we have this behavior note that as we decrease $\ell$ we have $\tilde{v}(k)$ and $\tilde{w}(k)$ supported over increasingly large $k$. For any given temperature the modes with high wavenumbers are themselves purer than modes with lower $k$. Intuitively, while the temperature $\beta$ is the same for each $k$ the number of excitations $\text{coth}(\beta\omega_k/2)$ decreases for increasing $k$.

Indeed, the shape of the mode (and not just the size) is important to determine the purity of the partial state of those modes when the total field state is pure. This points out that purity is extremely sensitive to the shape of the chosen mode: even a mode that occupies all the space in an optical cavity can be entangled with other modes---of different shape---also supported in the whole cavity.

We thus have very different qualitative features depending on what spatial profile is assigned to the 1+1 dimensional local modes. We now turn to the question of which of these behaviors we should expect in higher dimensions. Consider the 3D analogues of the B-spline functions defined in Eq.~\eqref{BMdef}, namely 
\begin{align}\label{B3D}
B_m^{(3D)}(x,y,z) \coloneqq \Pi_{(3D)}^{\ast m}(m\,x,m\,y,m\,z)    
\end{align}
where $\Pi_{(3D)}(x,y,z)$ is the characteristic function for the unit ball and the exponent $\ast m$ indicates repeated self-convolution. Eq.~\eqref{B3D} with\footnote{The case $m=1$ presents UV divergences associated with that particular shape.} \mbox{$m=2,3,4$} are profiles 7, 8 and 9 in Fig.~\ref{FigureZ}. The purities of these modes with a field mass $M$ are shown in Fig.~\ref{PurityFiguresSeries3}. As this figure shows, the purity of these 3+1 dimensional local modes is well defined in the massless limit and it is qualitatively similar to 1+1 dimensional modes considered in Fig.~\ref{PurityFiguresSeries2} (e.g, profiles 4, 5 and 6 in Fig.~\ref{FigureZ}). The behavior of these modes purities in the $M=0$ case are shown in Fig.~\ref{LimitPurity}b.

Same as we did before, we can set quasi-realistic atomic model by taking scales $\ell=53\text{ pm}$ and $T=4.2\text{ K}$. For these scales we have $k_\text{B}T\ell/\hbar c=9.7\times10^{-8}$ placing us on the far left of Fig.\ref{LimitPurity}a and Fig.\ref{LimitPurity}b. Note that the purity of these modes does not increase to unity as we decrease the ambient field temperature. Rather, the purity seems to max out at at most $P\approx0.95$. For these mode profiles, there appears to be a limit to how pure the local modes can be made by decreasing the ambient temperature.

This is a particular illustration of the claim in~\cite{ruep2021}: These modes are spatially localized and therefore are unavoidably entangled with other modes when the state of the field is the vacuum~\cite{Schlieder, Schlieder2}. However, this is certainly not what we would expect from a good model of an atomic probe. Indeed, atomic probes could be prepared in states as pure as $>0.999$ experimentally with technology now more than a decade old~\cite{TrappedIons}.  Furthermore, the amount of purity does depend strongly on the shape of the mode. Just specifying a localization scale is not enough to fully account for the mixedness behaviour of these modes. Indeed, as discussed in the previous subsection we could always find modes with a shape such that they are arbitrarily pure. What this points out is that  understanding these localized modes as a model for localized detectors (as for example atomic probes) is not a good idea.

\section{Threshold mixedness in UDW setup}\label{UDWperturbative}
    
  %\subsection{Field in a cavity}
We now turn to the task of quantitatively evaluating how much some initial mixedness in the probes can impact the amount of entanglement harvested. We are mainly interested in exploring the low-mixedness, small-coupling regime in more detail, where a perturbative analysis proves to be particularly useful. Each probe is again described by a localized quantum system coupled to a free scalar field in $D = n+1$ spacetime dimensions via an interaction action of the form in~\eqref{prototypeinteraction}:
\begin{equation}
    S_I = -\sum_{i\in\{\text{A},\text{B}\}} \lambda_i \int \dd^D x\sqrt{-g}\Lambda_i(\mathsf{x})\hat{\mu}_i(\tau_i)\phi(\mathsf{x}).
\end{equation}
Here, $\Lambda_i(\mathsf{x})$ are the two spacetime smearing functions that localize the support of the interaction region both in space and in time, and the sub-index $i=\text{A,B}$ labels each of the detectors. 

We then start the field-detectors system in a fully uncorrelated state,
\begin{equation}
    \hat{\rho}_0 = \hat{\rho}_\phi\otimes\hat{\rho}_{0, \textsc{ab}}
\end{equation}
with $\hat{\rho}_{0, \textsc{ab}}$ being the initial state of the probes. After the interaction, the state of the probes will be given by,
\begin{equation}
    \hat{\rho}_\textsc{ab} = \tr_{\phi}\left(\hat{U}\hat{\rho}_0\hat{U}^\dagger\right),
\end{equation}
where $U$ can be written in a way that is amenable to perturbation theory via the Dyson expansion:
\begin{align}
    \hat{U} =& \mathcal{T}\exp\left(-\dfrac{\ii}{\hbar}\int\dd t \hat{H}_I(t)\right) \nonumber \\
    =& \sum_{n=0}^\infty \left(\dfrac{-\ii}{\hbar}\right)^n \int_{-\infty}^{+\infty}\!\!\!\dd t_1 \dots \int_{-\infty}^{t_{n-1}}\!\!\!\dd t_n \hat{H}_I(t_1)\dots \hat{H}_I(t_n).
\end{align}
Here, we have implicitly picked a time coordinate $t$ which respect to which spacetime is being foliated, and the interaction Hamiltonian is given by Eq.~\eqref{HamiltonianT}, that is,
\begin{equation}
    \hat{H}_I(t) = \sum_{i\in\{\text{A},\text{B}\}}\lambda_i\int \dd^n\bm{x}\sqrt{-g}\Lambda_i(t, \bm{x})\hat{\mu}_i(\tau_i)\hat{\phi}(t, \bm{x}).
\end{equation}

The time evolution operator can be  expanded as,
\begin{equation}
    \hat{U} = \mathds{1} + \hat{U}^{(1)} + \hat{U}^{(2)}+\dots,
\end{equation}
where the term $\hat{U}^{(j)}$ collects all terms of order $\lambda^j$ in the coupling constants. This will lead to an expansion of the final state as
\begin{align}
    \hat{\rho} = \hat{\rho}_0 &+ \overbrace{ \hat{U}^{(1)}\hat{\rho}_0 + \hat{\rho}_0 \hat{U}^{(1)\dagger}}^{\propto \lambda} \nonumber\\&+ \underbrace{\hat{U}^{(1)}\hat{\rho}_0\hat{U}^{(1)\dagger} + \hat{U}^{(2)}\hat{\rho}_0 + \hat{\rho}_0 \hat{U}^{(2)\dagger}}_{\propto \lambda^2} +\dots.
\end{align}
After taking the trace over the field, assuming the initial state of the field to be a Gaussian (quasi-free) state with vanishing one-point function (which implies that all $n$-point functions of the field with $n$ odd vanish), the leading order contributions to the final state of the probes will be
\begin{align}
    \hat{\rho}_{\textsc{ab}}  =&\; \hat{\rho}_{0, \textsc{ab}}  \\
    &+ \tr_\phi\left(\hat{U}^{(1)}\hat{\rho}_0\hat{U}^{(1)\dagger} + \hat{U}^{(2)}\hat{\rho}_0 + \hat{\rho}_0 \hat{U}^{(2)\dagger}\right) + \mathcal{O}(\lambda^4).\nonumber
\end{align}
Now, by abbreviating the notation for the two-point function of the field as $\tr_\phi\left(\hat{\rho}_\phi \phi(\mathsf{x})\phi(\mathsf{x}')\right) \equiv W(\mathsf{x}, \mathsf{x}')$, denoting $\dd V\equiv \dd^D x\sqrt{-g}$, and implementing the time-ordering explicitly by inserting Heaviside functions in $t-t'$, we have
\begin{align}\label{finalstateperturbative}
    \hat{\rho}_{\textsc{ab}}  &= \hat{\rho}_{0, \textsc{ab}} + \sum_{i,j}\lambda_i\lambda_j \int\dd V\dd V'\Lambda_i(\mathsf{x})\Lambda_j(\mathsf{x}')\nonumber \\
    &\times\Big[\hat{\mu}_i(t)\hat{\rho}_{0, \textsc{ab}}\hat{\mu}_j(t')W(\mathsf{x}', \mathsf{x})\\
    &\qquad-\theta(t-t')\Big(\hat{\mu}_i(t)\hat{\mu}_j(t')\hat{\rho}_{0, \textsc{ab}}W(\mathsf{x}, \mathsf{x}')\nonumber \\ 
    &\quad\qquad + \hat{\rho}_{0, \textsc{ab}}\hat{\mu}_i(t')\hat{\mu}_j(t)W(\mathsf{x}', \mathsf{x})\Big)\Big] +\mathcal{O}(\lambda^4).\nonumber
\end{align}

We will then assign some initial mixedness for the two probes by taking as input initially uncorrelated thermal states. In other words, we will take,
\begin{equation}
    \hat{\rho}_{0, \textsc{ab}} = \hat{\rho}_{\textsc{a}}\otimes\hat{\rho}_{\textsc{b}},
\end{equation}
with each detector being described by,
\begin{equation}
    \hat{\rho}_{i} = \dfrac{1}{Z_i}e^{-\beta_i \hat{H}_{0,i}}.
\end{equation}
Here, $\hat{H}_{0,i}$ corresponds to the free Hamiltonian of the $i$-th detector which generates time translations with respect to its proper time, and $Z_i\coloneqq \tr e^{-\beta\hat{H}_{0,i}}$ is its partition function. For the case of a two-level UDW detector, this reduces to,
\begin{equation}
    \hat{\rho}^{\text{qubit}}_i = \dfrac{1}{1 + z_i}\big(\ket{0}\bra{0} + z_i\ket{1}\bra{1}\big),
\end{equation}
whereas for a harmonic-oscillator UDW detector, we have,
\begin{equation}
    \hat{\rho}^{\text{HO}}_i = (1-z_i)\sum_{n=0}^\infty z_i^n\ket{n}_i\bra{n}_i
\end{equation}
where $\{\ket{n}_i\}$ naturally corresponds to the eigenbasis of the free Hamiltonian of the $i$-th harmonic oscillator. In both cases, we have denoted the Boltzmann factor as,
\begin{equation}\label{defz}
    z_i = e^{-\beta_i \hbar\Omega_i}
\end{equation}
where $\beta_i$ is the inverse temperature of each probe, and $\Omega_i$ corresponds to its characteristic frequency (such that $\hbar\Omega$ is the energy gap between ground and excited states of the qubit detector, as well as the energy gap between any two consecutive eigenstates of a harmonic oscillator). It is clear that $z_i = 0$ corresponds to the pure case (detectors at $T=0$ or $\beta\rightarrow\infty$), and $z_i=1$ yields the maximally mixed state (with $T\rightarrow\infty$ or $\beta = 0$). 

This is a useful parametrization because it facilitates a perturbative expansion in the low-mixedness regime\footnote{Notice small mixedness may not mean small temperatures for human scales. For example, for a hydrogen atom ($\hbar\Omega=13.6$ eV) $z=0.001$ implies $T \sim 2.3\times 10^4$ K. For reference, this is about 4 times larger than the temperature at the surface of the Sun.}---which, as made explicit by the definition~\eqref{defz}, corresponds to the regime where the energy scale given by the temperature is much smaller than the detector's characteristic frequency. The purity of both the two-level system and the harmonic oscillator can be given simple expressions in terms of $z$: for the qubit case, we have,
\begin{equation}
    \Tr \hat{\rho}_{\text{qubit}}^2 = \dfrac{1 + z^2}{(1+z)^2} = 1 - 2z + \mathcal{O}(z^2),
\end{equation}
whereas for the harmonic oscillator,
\begin{equation}
    \Tr \hat{\rho}_{\text{HO}}^2 = \dfrac{1-z}{1+z} = 1 - 2z +\mathcal{O}(z^2).
\end{equation}
Therefore, in both cases, the dependence on the Boltzmann factor is exactly the same as long as we are only interested in a low-mixedness regime, where we only keep the leading order contributions in the Boltzmann factor. 

By looking at a low-mixedness regime and only keeping terms that are at most linear in $z_i$ on the initial state of the probes, both for the case with two qubit detectors and with two harmonic oscillators, the joint initial state of the detectors can be written as
\begin{align}
    \hat{\rho}_{0,\textsc{ab}} =& (1-z_{\textsc{a}} - z_{\textsc{b}})\ket{00}\bra{00} + z_{\textsc{a}}\ket{10}\bra{10} + z_{\textsc{b}}\ket{01}\bra{01} \nonumber\\
    &+ \mathcal{O}(z_{\textsc{a}}z_{\textsc{b}}),
\end{align}
where from now on we will abbreviate \mbox{$\ket{n}_{\textsc{a}}\otimes\ket{m}_{\textsc{b}} \equiv \ket{nm}$}. Carrying out a Dyson expansion~\eqref{finalstateperturbative} yields the following density matrix for the qubit (in the basis \mbox{$\{\ket{00}, \ket{01}, \ket{10}, \ket{11}\}$}):
\begin{widetext}
\begin{equation}\label{qubitstate}
    \hat{\rho}^{\text{qubit}}_{\textsc{ab}} = \begin{pmatrix}1 - (\mathcal{L}_{\textsc{aa}} + z_{\textsc{a}}) - (\mathcal{L}_{\textsc{bb}} + z_{\textsc{b}}) & 0 & 0 & \mathcal{M}^\ast \\ 0 & \mathcal{L}_{\textsc{bb}} + z_{\textsc{b}} & \mathcal{L}_{\textsc{ba}} & 0 \\ 0 & \mathcal{L}_{\textsc{ab}} & \mathcal{L}_{\textsc{aa}} + z_{\textsc{a}} & 0 \\ \mathcal{M} & 0 & 0 & 0 \end{pmatrix} +\mathcal{O}(z^2)+\mathcal{O}(z\lambda^2) + \mathcal{O}(\lambda^4).
\end{equation}
In the case of harmonic-oscillator detectors, there will be extra matrix elements coming from two excitations of each probe. In the basis $\{\ket{00}, \ket{01}, \ket{10}, \ket{11}, \ket{02}, \ket{20}\}$, the final state becomes,
\begin{equation}\label{HOstate}
    \hat{\rho}^{\text{HO}}_{\textsc{ab}} = \begin{pmatrix}1 - (\mathcal{L}_{\textsc{aa}} + z_{\textsc{a}}) - (\mathcal{L}_{\textsc{bb}} + z_{\textsc{b}}) & 0 & 0 & \mathcal{M}^\ast & \mathcal{K}_{\textsc{b}}^\ast & \mathcal{K}^\ast_{\textsc{a}}\\ 0 & \mathcal{L}_{\textsc{bb}} + z_{\textsc{b}} & \mathcal{L}_{\textsc{ba}} & 0 & 0 & 0\\ 0 & \mathcal{L}_{\textsc{ab}} & \mathcal{L}_{\textsc{aa}} + z_{\textsc{a}} & 0 & 0 & 0\\ \mathcal{M} & 0 & 0 & 0 & 0 & 0\\ \mathcal{K}_{\textsc{b}} & 0 & 0& 0& 0& 0\\ \mathcal{K}_{\textsc{a}} & 0 & 0& 0& 0& 0\end{pmatrix}+\mathcal{O}(z^2)+\mathcal{O}(z\lambda^2) + \mathcal{O}(\lambda^4).
\end{equation}
The various terms found above are given by\footnote{Note that to follow standard convention in entanglement harvesting we used $\mathcal{M}$ for the detectors' density matrix elements even though the same character was employed for the spacetime manifold in sections~\ref{twodetectormodels} and~\ref{RealisticMixedness}. We believe there is no notational clash as long as this is kept in mind.}
\begin{align}
    \mathcal{L}_{ij} &= \dfrac{\lambda_i\lambda_j}{\hbar^2}\int\dd V\dd V'e^{-\ii\Omega_i t}e^{\ii\Omega_j t'}\Lambda_i(\mathsf{x})\Lambda_j(\mathsf{x}')W(\mathsf{x}, \mathsf{x}'), \\
    \mathcal{M} &= -\dfrac{\lambda_{\textsc{a}}\lambda_{\textsc{b}}}{\hbar^2}\int\dd V\dd V'\Theta(t-t')W(\mathsf{x}, \mathsf{x}')[e^{\ii(\Omega_{\textsc{a}}t + \Omega_{\textsc{b}}t')}\Lambda_{\textsc{a}}(x)\Lambda_{\textsc{b}}(\mathsf{x}') + e^{\ii(\Omega_{\textsc{a}}t' + \Omega_{\textsc{b}}t)}\Lambda_{\textsc{a}}(\mathsf{x}')\Lambda_{\textsc{b}}(\mathsf{x})], \\
    \mathcal{K}_{i} &= -\dfrac{\lambda_i^2}{\hbar^2}\int\dd V\dd V'\Theta(t-t')e^{\ii\Omega_i(t+t')}\Lambda_i(\mathsf{x})\Lambda_i(\mathsf{x}')W(\mathsf{x}, \mathsf{x}').
\end{align}
\end{widetext}

It is thus clear that at leading order in both the coupling constant and the Boltzmann factor, the net effect of the nonzero temperature simply amounts to a correction to the populations of the first excited states of each detector. One can therefore trivially obtain the leading order corrections to entanglement harvesting due to mixedness by referencing the known results for initially pure states of the probes, and making the replacement \mbox{$\mathcal{L}_{ii} \rightarrow \mathcal{L}_{ii} + z_i$}.

A very common measure of entanglement in the context of entanglement harvesting is the \emph{negativity}, which is defined as the sum of the negative eigenvalues of the partial transpose of $\hat{\rho}_{\textsc{ab}}$ with respect to just one of its subsystems. Negativity is well-defined for mixed states of arbitrary-dimensional Hilbert spaces and therefore allows for a direct comparison between different variants of the UDW model (qubit, harmonic-oscillator detector, etc.). %For bipartite systems of two qubits and for Gaussian states of two single-mode harmonic oscillators (which fortunately happen to be the cases of interest here\footnote{Rigorously speaking, the perturbative calculation presented in this subsection is not explicitly using the Gaussian formalism of quantum mechanics and the Dyson expansion does not preserve gaussianity. However, since the Hamiltonian is quadratic in the quadrature operators of field and detectors, an exact calculation \emph{would} lead to a final state of the probes that is indeed Gaussian. For consistency, one should then expect that such (nonperturbative) solution will yield to the same leading-order dependence on the coupling and mixedness as we are finding here, once expanded to the same order in the Boltzmann factor and coupling constant.}),nonzero negativity is both a sufficient \emph{and} necessary condition for the joint state to be nonseparable\cite{}. 
For both states~\eqref{qubitstate} and~\eqref{HOstate}, it turns out that the partial transpose can have at most one negative eigenvalue at leading order in $\lambda^2$, given by~\cite{Pozas-Kerstjens:2015},
\begin{equation}
        E_{1}\!=\!\frac{1}{2}\left[P_{\textsc{a}}\!+\!P_{\textsc{b}}\!-\!\sqrt{\left(P_{\textsc{a}}\!-\!P_{\textsc{b}}\right)^{2}\!+\!4|\mathcal{M}|^{2}}\right]+\mathcal{O}(z^2,z\lambda^2, \lambda^{4}),
    \end{equation}
where here $P_i \coloneqq \mathcal{L}_{ii} + z_i$. The negativity can then be written as \mbox{$\mathcal{N}=\mathcal{N}^{(2)}+\mathcal{O}(z^2,z\lambda^2, \lambda^{4})$} where,
\begin{equation}
    \mathcal{N}^{(2)} = \text{max}(-E_1, 0).
\end{equation}
If the Boltzmann factor of both detectors is the same (that is, \mbox{$z_{\textsc{a}} = z_{\textsc{b}} \equiv z$}), this expression is further simplified to,
\begin{equation}\label{negativitywithmixedness}
    \mathcal{N}^{(2)} = \max\left(\mathcal{N}^{(2)}_{z=0} - z, 0\right),
\end{equation}
where $ \mathcal{N}^{(2)}_{z=0}$ corresponds to the negativity computed at second order in the coupling for the case of perfectly pure initial probes. This highlights the fact that the presence of any initial mixedness systematically decreases the entanglement harvested. One can clearly identify the existence of a threshold mixedness, $z_c = \mathcal{N}^{(2)}_{z=0}$, above which one no longer has any entanglement in the final state of the probes at lowest order in perturbation theory. 

The result for the negativity simplifies even further when we consider the particular case with two inertial detectors in flat spacetime with identical smearing functions and the same energy gap. In that configuration, we will have \mbox{$\mathcal{L}_{\textsc{aa}} = \mathcal{L}_{\textsc{bb}}\equiv \mathcal{L}_{ii}$}, and therefore the negativity reduces to,
\begin{equation}
    \mathcal{N}^{(2)} = \max\left(\abs{\mathcal{M}} - \mathcal{L}_{ii} - z, 0\right).
\end{equation}
This is a well known expression that has often been described as a competition between the non-local term $\abs{\mathcal{M}}$ and the local noise term $\mathcal{L}_{ii}$~\cite{Reznik1}. We thus see, very intuitively, that entanglement harvesting is only possible at this level in perturbation theory if the nonlocal term $\abs{\mathcal{M}}$, which involves the two-point function of the field evaluated along the trajectories of both detectors, overcomes the local ``noise'' terms. For the case of initially pure probes, the local noise only includes the vacuum excitation probability term (given by $\mathcal{L}_{ii}$), but when the initial state is already thermally populated due to nonzero temperature of the detector, one also has to take the Boltzmann factor $z$ into account. 

In summary, we have shown that, in the usual framework of UDW-like particle detector models commonly employed in RQI, initial mixedness of the probes does indeed hinder the ability to harvest entanglement perturbatively. This vindicates the qualitative claim made in~\cite{ruep2021}, where probe mixedness was identified as the culprit for their detector model not being able to extract entanglement from the vacuum of a quantum field for low enough coupling. Here, we have provided evidence that this fact is indeed a generic feature of initially mixed probes, that can also be made manifest in UDW-like detectors such as the qubit or the harmonic oscillator. 

We also see that, within the context of UDW-like detector models, the threshold mixedness that allows for harvesting (or equivalently, the critical coupling strength $\lambda_c$ that allows for harvesting at a given mixedness) can be readily computed at lowest order with a perturbative calculation. Since $\mathcal{N}^{(2)}_{z=0}\propto\lambda^2$, for any sufficiently high purity $1-2z$, the threshold coupling strength (i.e., the minimum value for the coupling constant necessary in order for entanglement to be harvested at lowest order) scales as \mbox{$\lambda_c\propto\sqrt{z}$}. In particular, if the mixedness of the probe comes from the fact that it is at a small but finite temperature then $\sqrt{z} = \exp\left(-\hbar\Omega/2k_{\textsc{b}}T\right)$. This means that the critical coupling strength is exponentially suppressed with $1/T$, showing that the well-known results on perturbative entanglement harvesting are effectively protected from the problem pointed out in~\cite{ruep2021} due to the mixedness of the probe, for sufficiently low temperatures.

\section{Conclusion}
In this paper we have overviewed the differences between the Fewster-Verch approach to modeling the measurements of quantum fields and the usual Unruh-DeWitt particle detector models commonly employed in relativistic quantum information and QFT in curved space. We have conducted this review in order to respond to a recent paper \cite{ruep2021} which has claimed (using a development of the Fewster-Verch approach) that entanglement harvesting is not possible with weak coupling strengths. 

In brief, their argument is as follows. As a consequence of Reeh-Schlieder theorem any localized mode of a quantum field in a Reeh-Schlieder state will be mixed. If our local probes are ultimately built out of local modes of quantum fields, then they will necessarily inherit this mixedness. If our local probes begin in mixed states, then there is some finite interaction strength which is required to get them entangled. This is because for any mixed quantum states $\hat\rho_A$ and $\hat\rho_B$, there is an $\epsilon>0$-sized ball of separable states which is centered at $\hat\rho_A\otimes\hat\rho_B$. Thus, \cite{ruep2021} proves that there is a critical coupling strength below which entanglement harvesting is not possible. While the above argument is valid, once put into context its conclusion and further consequences are seen to be of little consequence to experimentally realistic scenarios. 

We began our review of the difference between the Fewster-Verch framework and Unruh-DeWitt model, in Sec.~\ref{RQMProblem}, by discussing the quantum measurement problem in both nonrelativistic and relativistic contexts. In particular we have discussed the role that a Heisenberg cut plays in the nonrelativistic setting in order to motivate an analogous \textit{relativistic cut} in a relativistic setting. By our reckoning one of the major differences between  the Fewster-Verch framework and the particle detector framework stems from how they handle this relativistic cut. The Fewster-Verch approach measures a quantum field with another quantum field. This begs the question of how that field is itself to be measured. The Unruh-DeWitt model provides a justifiable stopping point for this regression (at least for modeling purposes).% Following this, in Sec.~\ref{twodetectormodels} we have provided the mathematical details of these two approaches.

Having established this background, we put into context the observation in~\cite{ruep2021} that the reduced state of any local mode will be mixed due to the Reeh-Schlieder theorem. In addition to this Reeh-Schlieder mixedness we have noted that realistic detectors will be mixed simply due to them having a finite temperature.  We have provided a quantitative analysis of how the total mixedness of local modes of a free field depends on both the size and shape of the mode as well as the ambient temperature of the field. In particular, we have shown that, for any fixed size of the local mode and temperature of the field, it is always possible to make the local mode as pure as we like by changing its shape. Thus, while it is true that local modes are necessarily mixed, there is no non-trivial lower bound on the mixedness in terms of either the region's size or the ambient field temperature. 

Additionally we have investigated for a local mode with a fixed shape the effects of the mode's size and the ambient temperature on the mode's purity. As we have discussed, the dependence of the local mode's purity on its shape, size, and the ambient temperature all indicate that local field modes of a free theory are not good models of the localized probes which are experimentally available. Thus, to achieve an adequate model of realistic probe it is apparently necessary to forgo local modes of free theories and move on to use bound states of interacting theories. While this is possible in principle within the Fewster-Verch framework, for all practical purposes this is currently arguably unfeasible. By contrast, the UDW model stands as a workable alternative to modeling our local probes as quantum fields themselves. Indeed, a measurement theory for quantum fields~\cite{Polo} has recently been put forward which: a) is compatible with relativity, b) provides an update rule as well as definite single-shot outcomes, and (most crucially) c) has a clear and direct connection with laboratory experiments.

Finally, in Sec.~\ref{UDWperturbative} we investigated the claim in~\cite{ruep2021} that the initial mixedness of detectors adversely affects their ability to harvest entanglement with weak coupling strengths. In particular,~\cite{ruep2021} has proven that for any fixed non-zero level of probe mixedness there will be a critical coupling threshold below which entanglement harvesting is not possible. Thus, at a fixed level of mixedness we cannot study entanglement harvesting perturbatively in the coupling strength. In response we have conducted a perturbative analysis of entanglement harvesting in \textit{both} the coupling strength and the initial probe mixedness. As we have shown when the initial mixedness and coupling strength satisfy a simple inequality entanglement harvesting is indeed possible.

\section{Acknowledgements}
The authors thank Christopher Fewster and Maximilian Ruep for stimulating discussions on the measurement problem in QFT, as well as Jos\'{e} Polo-G\'{o}mez for his helpful feedback. E.M-M acknowledges the support of the NSERC Discovery program as well as his Ontario Early Researcher Award. B.S.L.T. acknowledges support from the Mike and Ophelia Lazaridis Fellowship. Research at Perimeter Institute is supported in part by the Government of Canada through the Department of Innovation, Science and Economic Development Canada and by the Province of Ontario through the Ministry of Colleges and Universities.

\appendix

\section{Gaussian quantum mechanics}\label{GQM}
    
    The quantum state $\hat{\rho}$ of a collection of bosonic particles can be described in terms of a quasi-probability distribution on phase space, given by the so-called \emph{Wigner function}. For a system of particles described by $n$ positions and $n$ momenta, the Wigner function $W_{\hat{\rho}}(\bm{q}, \bm{p})$ representing the state $\hat{\rho}$ can be written as,
    \begin{equation}\label{defWignerFunction}
    W_{\hat{\rho}}(\bm{q}, \bm{p}) = \dfrac{1}{(2\pi\hbar)^n}\int \dd^n x \,\,e^{-\ii \bm{p}\cdot \bm{x}/\hbar}\bra{\bm{q} + \dfrac{\bm{x}}{2}}\hat{\rho}\ket{\bm{q} - \dfrac{\bm{x}}{2}},
    \end{equation}
    where $\ket{\bm{q} \pm \bm{x}/2}$ correspond to eigenstates of the position operator $\hat{\bm{X}}$ with eigenvalues $\bm{q} \pm \frac{\bm{x}}{2}$ respectively, and \mbox{$\bm{p} \cdot\bm{x} \equiv p_i x^i$} is the standard inner product in $\mathbb{R}^n$. 
    
    The Wigner function can also be written in terms of the momentum-basis matrix elements of $\hat{\rho}$, as,
    \begin{equation}\label{defWignerFunction2}
     W_{\hat{\rho}}(\bm{q}, \bm{p}) = \dfrac{1}{(2\pi\hbar)^n}\int \dd^n k\,e^{\ii \bm{k}\cdot \bm{q}/\hbar}\Big\langle\bm{p} + \dfrac{\bm{k}}{2}\Big|\, \hat{\rho}\,\Big|\bm{p} - \dfrac{\bm{k}}{2}\Big\rangle
    \end{equation}
    where now $\ket{\bm{p} \pm \bm{k}/2}$ are momentum eigenstates.
    Finally, the expression for $ W_{\hat{\rho}}(\bm{q}, \bm{p})$ can also be put in a form that leaves position and momentum explicitly on equal footing, given by,
    \begin{equation}\label{WignerFunctionFinalPhaseSpace}
    W_{\hat{\rho}}(\bm{\xi}) = \dfrac{1}{(2\pi\hbar)^{2n}}\int \dd^{2n} \xi'\,\,e^{\frac{\ii}{\hbar} \xi^\alpha\Tilde{\Omega}_{\alpha\beta}\xi'^\beta}\expval{e^{\frac{\ii}{\hbar}\xi'^\alpha\Tilde{\Omega}_{\alpha\beta}\hat{\Xi}^\beta}}_{\hat{\rho}}.
    \end{equation}
    where now $\dd^{2n} \xi' = \dd^n q' \dd^n p'$ is a full phase-space integral, and we have defined the $2n$-dimensional phase-space vector,
    \begin{equation}
    \bm{\xi} = (q^1, p_1, \dots, q^n, p_n)^\intercal
    \end{equation}
    as well as the $2n\times2n$ symplectic matrix,
    \begin{equation}\label{sympform}
    (\tilde{\Omega}_{\alpha\beta}) = \bigoplus_{i = 1}^n \begin{pmatrix}0 & -1 \\ 1 & 0 \end{pmatrix}
    \end{equation}
    and the generalized phase-space operator,
    \begin{equation}
    \hat{\bm{\Xi}} = (\hat{X}^1, \hat{P}_1, \dots, \hat{X}^n, \hat{P}_n)^\intercal.
    \end{equation}
     Eq.~\eqref{WignerFunctionFinalPhaseSpace} thus shows that $\bm{q}$ and $\bm{p}$ play a completely symmetric role in the definition of $W_{\hat{\rho}}(\bm{q}, \bm{p})$. 
    
    The canonical commutation relations,
    \begin{equation}
    \comm{\hat{X}^i}{\hat{P}_j} = \ii \hbar \delta^{i}_{j}\mathds{1},
    \end{equation}
    can be written in terms of $\hat{\bm{\Xi}}$ as,
    \begin{equation}
        \comm{\hat{\Xi}^\alpha}{\hat{\Xi}^\beta} = \ii \hbar \Tilde{\Omega}^{\alpha\beta}\mathds{1},
    \end{equation}
    where we have,
    \begin{equation}\label{inversesympform}
    (\tilde{\Omega}^{\alpha\beta}) = \bigoplus_{i = 1}^n \begin{pmatrix}0 & 1 \\ -1 & 0 \end{pmatrix}
    \end{equation}
    which is the inverse of~\eqref{sympform}, so that $\Tilde{\Omega}^{\alpha\lambda}\Tilde{\Omega}_{\lambda\beta} = \delta^{\alpha}_{\beta}$.
    
    The Wigner function corresponds to a real, unit-normalized function, whose marginals on $\bm{q}$ and $\bm{p}$ correspond precisely to the probability densities on $\bm{p}$ and $\bm{q}$ respectively. The Wigner function is also, in a way, analogous to a probability distribution when it comes to computing expectation values of observables: for any operator $\hat{A}$ corresponding to the Weyl-quantized~\cite{Weyl, Weyl2} version of a phase space observable $A(\bm{\xi})$, one can show that, 
    \begin{equation}
         \expval{\hat{A}}_{\hat{\rho}} = \Tr \left(\hat{\rho}\hat{A}\right) = \int \dd^{2n} \xi\,W_{\hat{\rho}}(\bm{\xi})A(\bm{\xi}).
    \end{equation}
    
    However, the Wigner function generally cannot be seen as a legitimate probability distribution on phase space, as it fails to be positive-semidefinite. In other words, for a generic state $\hat{\rho}$, there will be regions on phase space where $W_{\hat{\rho}}(\bm{q}, \bm{p})$ attains negative values. The Wigner function is therefore most appropriately described as a \emph{quasi}-probability distribution. For pure states, the only states whose Wigner functions are positive-definite are \emph{Gaussian states}, which---as the name suggests---are represented by Gaussian functions on the phase space variables:
    \begin{equation}\label{gaussianstates}
    W_{\hat{\rho}}(\bm{\xi}) = \dfrac{1}{(\pi\hbar)^n\sqrt{\det(\Sigma)}} \exp\left[-\left(\bm{\xi} - \bm{\xi}_0\right)^\intercal\Sigma^{-1}\left(\bm{\xi} - \bm{\xi}_0\right)\right].
    \end{equation}
    It is easy to check that,
    \begin{align}
    \xi_0^\mu =& \expval{\hat{\Xi}^\mu}_{\hat{\rho}}, \\
    \Sigma^{\mu\nu} =& \expval{\hat{\Xi}^{\mu}\hat{\Xi}^{\nu} + \hat{\Xi}^{\nu}\hat{\Xi}^{\mu}}_{\hat{\rho}} - 2\xi_0^\mu \xi_0^\nu\nonumber \\
     =& \expval{\left(\hat{\Xi}^{\mu} - \xi_0^\mu\right)\left(\hat{\Xi}^{\nu} - \xi_0^\nu\right)}_{\hat{\rho}} \nonumber \\ &+ \expval{\left(\hat{\Xi}^{\nu} - \xi_0^\nu\right)\left(\hat{\Xi}^{\mu} - \xi_0^\mu\right)}_{\hat{\rho}},\label{covmatrix}
    \end{align}
and through usual tricks in Gaussian integrals, one can see that the higher-order statistical moments of the distribution are fully determined by the first moment $\xi_0^\mu$ and the covariance matrix $\Sigma^{\mu\nu}$.

When the initial state of the system is Gaussian and the underlying dynamics preserves Gaussianity, one is able to fully characterize the state at all times by just keeping track of the first moments and the covariance matrix, thus rendering the description of the dynamics effectively finite-dimensional. Unitary transformations that preserve Gaussianity are generated by Hamiltonians that are at most quadratic on the phase space operators (also called quadratures) $\hat{\Xi}^\mu$: therefore, any such unitary $\hat{U}$ can be written as,
\begin{align}
    \hat{U}(t) =& \exp\left(-\dfrac{\ii}{\hbar}\hat{H}t\right), \\
    \hat{H} =& \dfrac{1}{2}\hat{\Xi}^\mu F_{\mu\nu}\hat{\Xi}^\nu + \alpha_{\mu}\hat{\Xi}^\mu \label{quadratichamiltonian}
\end{align}
where $\bm{F}$ is a $2n \times 2n$ Hermitian matrix, and $\bm{\alpha}$ is a real-valued vector\footnote{We are technically assuming the Hamiltonian to be time-independent, but of course one could let $\bm{F}$ and $\bm{\alpha}$ have some explicit time dependence; the only thing that would change would be that the time evolution operator would be written in terms of a time-ordered exponential, instead of the simple exponential.}. This will lead to an evolution of the quadratures given by,
\begin{equation}
    \hat{\bm{\Xi}}(t) = \bm{S}(t)\hat{\bm{\Xi}} + \bm{d}(t)\mathds{1},
\end{equation}
where $\bm{S}(t)$ is a matrix and $\bm{d}(t)$ is a vector, given respectively by,
\begin{align}
    \bm{S}(t) = \exp\left(\Tilde{\Omega}^{-1}\bm{F}t\right), \\
    \bm{d}(t) = \dfrac{\bm{S}(t) - \mathds{1}}{\Tilde{\Omega}^{-1}\bm{F}}\Tilde{\Omega}^{-1}\bm{\alpha}
\end{align}
and $\Tilde{\Omega}^{-1}$ is the inverse of the symplectic form, given in~\eqref{inversesympform}. In summary, Gaussian unitaries generate linear-affine transformations on the quadratures. The linear part of the Hamiltonian~\eqref{quadratichamiltonian} generates translations (displacements) on phase space, and the quadratic part generates rotations as well as other operations known as one- and two-mode squeezing and beam-splitting~\cite{gaussianquantuminfo}. It is straightforward to show that the first moments and covariance matrix transform under Gaussian unitaries as,
\begin{align}
    \bm{\xi}_0 \rightarrow& \bm{S}(t)\bm{\xi}_0 + \bm{d}(t), \\
    \Sigma \rightarrow& \bm{S}\Sigma \bm{S}^\intercal
\end{align}
and that $\bm{S}(t)$ preserves the symplectic form: $\bm{S}(t)\Tilde{\Omega}\bm{S}^\intercal(t) = \Tilde{\Omega}$.
More generally, transformations on the quantum state (not necessarily unitary) that preserve Gaussianity are called Gaussian channels. It can be shown that all Gaussian channels can be dilated into Gaussian unitaries that act on a higher-dimensional phase space, in clear analogy to the general Stinespring dilation theorem for completely positive trace-preserving maps.

Measures of mixedness and entanglement of a Gaussian state are encoded solely in the covariance matrix $\Sigma$. %Notice that through a phase-space translation, implemented by displacement operators which are generated by Hamiltonians that are linear in the quadratures, we can always set $X^\mu = 0$, in which case the covariance matrix reduces to a collection of expectation values of the anticommutator of the quadrature operators. Since displacements are ``single-mode'' operations (they act on each mode separately), they do not affect measures of entanglement. 
The covariance matrix is symmetric and positive-definite, and Williamson's theorem~\cite{Williamson} guarantees that there is a symplectic transformation---i.e., a linear map $\bm{\xi}\rightarrow S\bm{\xi}$ that preserves the symplectic form, $S\Tilde{\Omega}S^\intercal = \Tilde{\Omega}$---that takes $\Sigma$ to diagonal form. This means that one can always write $\Sigma_D = S\Sigma S^\intercal$,
where $\Sigma_D$ is of the form,
\begin{equation}
    \Sigma_D = \bigoplus_{i = 1}^n \begin{pmatrix}\nu_i & 0 \\ 0 & \nu_i \end{pmatrix}.
\end{equation}
The set $\{\nu_i\}$ is called the \emph{symplectic spectrum} of the covariance matrix $\Sigma$, and the elements of the symplectic spectrum are called \emph{symplectic eigenvalues}. It can be shown that for any Gaussian state, one must have $\nu_i \geq 1$---which is a manifestation of the uncertainty principle, intuitively reinforcing the existence of a ``minimum area'' in phase space by preventing the covariances to be too small.

Written in the coordinates on phase space that symplectically diagonalize the covariance matrix (which are analogous to a basis of normal modes), the Wigner function becomes a product of single-mode Wigner functions. Taking the first moments to vanish (which can always be achieved by a unitary, single-mode phase space displacement) this means that we can write,
\begin{equation}
    W_{\hat{\rho}}(\bm{\xi}) = \prod_{j=1}^n W_{j}(\bm{\xi}_j)
\end{equation}
where $\bm{\xi}_j = (q^j, p_j)$, and we have,
\begin{equation}
    W_{j}(\bm{\xi}_j) = \dfrac{1}{\nu_j\pi}\exp\left(-\dfrac{(q_j^2 + p_j^2)}{\nu_j}\right).
\end{equation}
From this we can see that, roughly speaking, the simplectic eigenvalue can be interpreted as an occupation number of a normal mode in the state $\hat{\rho}$. More generally, other measures of mixedness---such as the Von Neumann entropy---are monotonic functions of $\nu_j$, and therefore the symplectic eigenvalues provide a very intuitive and useful parametrization for the mixedness of Gaussian states.

%\twocolumngrid

\bibliography{references}

\end{document}